\documentclass[manuscript]{aastex}

\usepackage{longtable}
\usepackage{amsbsy}
\usepackage{wasysym}
\usepackage{color}

\shorttitle{JVLA survey of the Serpens region}
\shortauthors{Ortiz-Le\'on et al.}

\begin{document}

\title{The Gould's Belt Very Large Array Survey II:\\ The Serpens region}

\author{ Gisela N. Ortiz-Le\'on\altaffilmark{1}, 
Laurent Loinard\altaffilmark{1,2}, 
Amy J. Mioduszewski\altaffilmark{3}, 
Sergio A. Dzib\altaffilmark{2},
Luis F. Rodr\'iguez\altaffilmark{1,4},
Gerardo Pech\altaffilmark{1},
Juana L. Rivera\altaffilmark{1},
Rosa M. Torres\altaffilmark{5},
Andrew F. Boden\altaffilmark{6},
Lee Hartmann\altaffilmark{7},
Neal J. Evans II\altaffilmark{8},
Cesar Brice\~no\altaffilmark{9,10},
John Tobin\altaffilmark{11,12},
Marina A. Kounkel\altaffilmark{7},
and Rosa A. Gonz\'alez-L\'opezlira\altaffilmark{1}
}

\email{g.ortiz@crya.unam.mx}

\altaffiltext{1}{Centro de Radioastronom\'ia y Astrof\'isica, 
Universidad Nacional Aut\'onoma de Mexico,
Morelia 58089, Mexico}
\altaffiltext{2}{Max Planck Institut f\"ur Radioastronomie, Auf dem H\"ugel 69, D-53121 Bonn, Germany}
\altaffiltext{3}{National Radio Astronomy Observatory, Domenici Science Operations Center, 1003 
Lopezville Road, Socorro, NM 87801, USA}
\altaffiltext{4}{King Abdulaziz University, P.O. Box 80203, Jeddah 21589, Saudi Arabia}
\altaffiltext{5} {Instituto de Astronom\'ia y Meteorolog\'ia, Universidad de Guadalajara, Av. 
Vallarta 2602, 
Col. Arcos Vallarta, 44130, Guadalajara, Jalisco, M\'exico. }
\altaffiltext{6}{Division of Physics, Math and Astronomy, California Institute of Technology, 1200 East
California Boulevard, Pasadena, CA 91125, USA} 
\altaffiltext{7}{Department of Astronomy, University of Michigan, 500 Church Street, 
Ann Arbor, MI 48105, 
USA}
\altaffiltext{8}{Department of Astronomy, The University of Texas at Austin, 
2515 Speedway, Stop C1400, Austin, TX 78712-1205, USA}
\altaffiltext{9}{Centro de Investigaciones de Astronom\'ia, M\'erida 5101-A, Venezuela}
\altaffiltext{10}{Cerro Tololo Interamerican Observatory, Casilla 603, La Serena, Chile}
\altaffiltext{11}{National Radio Astronomy Observatory, Charlottesville, VA 22903, USA}
\altaffiltext{12}{Hubble Fellow}

\begin{abstract}
We present deep ($\sim 17~\mu$Jy) radio continuum observations of the Serpens molecular cloud, 
the Serpens south cluster, and the W40 region obtained using the Very Large 
Array in its A configuration. We detect a total of 146 sources, 29 of which are 
young stellar objects (YSOs), 2 are BV stars and 5 more are associated
with phenomena related to YSOs.
Based on their radio variability and spectral index, we propose that about 16
of the remaining 110 unclassified sources are also YSOs. For approximately 65\%
of the known YSOs detected here as radio sources, the emission is most likely
non-thermal, and related to stellar coronal activity. As also recently observed 
in Ophiuchus, our sample of YSOs with X-ray counterparts lies below the fiducial 
G\"udel \& Benz relation. Finally, we analyze the proper motions of 9 sources 
in the W40 region. This allows us to better  constrain the membership of the radio sources 
in the region.
\end{abstract}

\keywords{astrometry - magnetic fields -  radiation mechanisms: non-thermal -
radio continuum: stars - techniques: interferometric}

\section{Introduction}\label{sec:intro}

Radio continuum observations toward star-forming regions are  relevant 
because they provide insights into thermal and non-thermal emission in 
young stellar objects (YSOs), stellar 
coronal activity of YSOs, and magnetic fields.
Different processes are invoked to explain the origin of the radio emission in 
these different kinds of objects.
Embedded Class I protostars have most often been detected
as thermal bremsstrahlung sources, and this emission is predominantly due to collimated
thermal winds or jets. In the case of more massive stars, the radio
emission can also originate from optically thick or thin compact HII
regions (Hughes 1988; Estalella et al.\ 1991; G\'omez et al.\ 2000), 
or from ionized winds (Felli et al.\ 1998). Non-thermal (gyrosynchrotron)
emission has also been detected in a number of sources. This 
mechanism produces radiation characterized by high brightness 
temperature, high variability, and often a negative spectral index and
some level of circular polarization (e.g.\ Hughes 1991; Hughes et al.\ 1995; 
Garay et al.\ 1996). This  non-thermal radio emission is generally present in more 
evolved YSOs (Class III sources), but it has also been detected in a number of 
Class II and even in a few Class I sources (e.g. Forbrich et al.\ 2007; 
Dzib et al.\ 2010; Deller et al.\ 2013).
We note that very little has been done on the characterization of 
the Serpens, Serpens South and W40 regions at radio wavelengths.

The Aquila rift/Serpens complex is one of the clouds selected for observations 
as part of \emph{The Gould's Belt Distance Survey}, which is a large project designed 
to determine accurate distances to stars in the most often studied
star-forming sites (Loinard 2013).
In this paper we report on new sensitive and high angular resolution radio 
observations of the Serpens, Serpens South and W40 regions. This paper is the second 
(after that by Dzib et al. 2013 which dealt with the 
Ophiuchus region) in a series that will focus on the analysis of the radio 
emission from YSOs in the star-forming regions of the Gould's Belt using the
Karl G. Jansky Very Large Array (JVLA). 
The observations cover large fields of view of the 
three regions: 900, 290 and 280 square arcminutes in Serpens, Serpens South
and W40, respectively, i.e., considerably larger than previous observations
carried out at radio wavelengths toward these regions.

The Serpens molecular cloud has been studied via multi-wavelength 
observations ever since it was recognized as  an active star-forming region 
by Strom et al.\ (1974).  The cloud belongs  to a larger complex of local optically 
dark molecular clouds called the Aquila Rift, which, in turn, is part of the Gould's Belt
(e.g., Dame et al.\ 1987; Perrot \& Grenier 2003). The Serpens cloud shows large scale 
irregular dark structures  in optical images and several nebulae can be distinguished. 
One of the most prominent is the Serpens nebula, which is
illuminated by the pre-main sequence (PMS) star SVS 2
(Strom et al.\ 1974, 1976, Worden \& Grasdalen 1974, King et al.\ 1983, 
Warren-Smith et al.\ 1987, G\'omez de Castro et al.\ 1988). The region, about 
6$'$ across centered on the Serpens nebula, is known as the Serpens cloud core, and 
was described in early observations by Loren et al.\ (1979) as a nearly circular, 
high density formaldehyde (H$_2$CO) core.  A more elongated structure extends in a  
North-West/South-East direction as seen, for instance, on maps of far-IR and NH$_3$ 
emission (e.g., Torrelles et al.\ 1989). The Serpens core is populated by 
more than $\sim 300$ objects found in many different evolutionary stages and
coexisting within the central $\sim 0.5 - 0.7$ pc of the core (Eiroa et al.\
2008 and references therein). 

Near- and mid-infrared (IR) observations have identified a large number of
Class II PMS stars, flat-spectrum sources and
Class I protostars embedded in the cloud core (Eiroa \& Casali 1992; 
Sogawa et al.\ 1997; Giovannetti et al.\ 1998; Kaas 1999; Kaas et al.\ 2004). 
Class 0 protostars and protostellar condensations have also been found
by means of submillimeter, millimeter, and far-IR observations (Casali et al.\ 1993; 
Hurt \& Barsony 1996; Testi \& Sargent 1998). In addition, some of these 
YSOs have X-ray counterparts (Preibisch 1998, 
2003, 2004; Giardino et al.\ 2007). 
\emph{Spitzer}/IRAC and MIPS observations extend to a larger region of
$\sim 6-9$ pc (Harvey et al.\ 2006, 2007a, 2007b; Oliveira et al.\ 2010). 
Briefly, these most recent works identified at least two main centers of star 
formation, which were named Clusters A and B. Cluster A is the already described region 
referred to as the Serpens Core, while Cluster B, $\sim 35'$ to the south of Cluster 
A and also referred to as Serpens G3-G6, was not observed by us.
At radio wavelengths, the Serpens Core 
has been observed by Eiroa et al.\ (2005) with the Very Large Array. 
A total of 22 radio continuum sources were detected, 16 of which were proposed 
to be associated with Class 0, Class I, flat spectrum, and Class II YSOs of the core. 
Scaife et al.\ (2012) carried out deep radio continuum observations at 1.8 cm with the 
Arcminute Microkelvin Imager Large Array (providing an angular resolution of 
$\sim 30$ arcsec) of the 19 protostellar cores 
reported in the \emph{Spitzer} catalog of Dunham et al.\ (2008).
They detected 8 radio sources, 6 possibly associated with deeply embedded young
stellar objects.

At an angular distance of $\sim 3^{\circ}$ to the south of the center of the 
Serpens core lies the star forming region known as W40. Assuming a distance of 
415 pc (see discussion below), this corresponds to a spatial distance of
$\sim$22 pc. Three main 
components are recognized in this region. First, the cold molecular cloud 
G28.8$+$3.5 (Goss \& Shaver 1970)
with an extent of $\sim 1$ deg, centered around the smaller ($\sim 20'$) dense 
molecular core, TGU 279$-$P7  (Dobashi et al.\ 2005). Second, there is a 
blister HII region (W40) of diameter $\sim 6'$ centered on J2000 coordinates 
$18^{\rm h}\,30^{\rm m}\,29^{\rm s}$, $-2^{\circ} \,05\rlap.'4$, and adjacent to 
the molecular cloud (Westerhout 1958). 
Finally, the W40 region hosts an embedded OB star cluster that is the 
primary excitation source for the W40 HII region (Smith et al.\ 1985).
There is evidence for  on-going star formation from dense 
molecular material in this region, as suggested by the detection of
an IRAS source, a cold ammonia core, and a
number of millimeter-wave sources  associated with the
cluster (Molinari et al.\ 1996; Maury et al.\ 2011). 

The stellar population of W40 has also been investigated through observations 
at different wavelengths.  Crutcher \& Chu (1982) and Smith et al.\ (1985) 
found seven IR bright sources with optical counterparts behind 
9$-$10 mag of visual extinction within the central $3'$ of W40. 
According to the spectral energy distributions (SEDs)  from the IR through 
the millimeter obtained by Smith et al.\ (1985) and Vallee \& MacLeod (1994),
most of these bright sources are surrounded by significant amounts of
circumstellar material.
A cluster of near-IR sources is detected in the Two Micron All Sky Survey (2MASS) 
images within the central $5'$ (Smith et al.\ 1985). Recently, Rodr\'iguez et al.\ 
(2010) observed the W40 region at 3.6 cm with the Very Large Array (VLA). 
They found a cluster of 20 compact sources in the central portion of the W40 
IR cluster, many of which correspond to the known IR sources. The W40 cluster 
has also been observed with the Chandra X-Ray Observatory. 
These observations reveal approximately 200 sources associated with the 
cluster, the majority of which are thought to be low-mass
YSOs (Kuhn et al.\ 2010). More recently, Shuping et al.\ (2012) determined 
the spectral classification and SEDs of the brightest members of the central 
stellar cluster in the W40 region. They identified four main sequence 
OB stars, two Herbig AeBe stars, and two low-mass YSOs (Class II).

Very close to W40 in projection on the plane of the sky lies an embedded cluster
of IR sources referred to as Serpens South. This cluster was recently 
discovered by Gutermuth et al.\ (2008) from \emph{Spitzer} observations of the Aquila 
rift region. They identified 54 sources classified as either Class I or flat SED
and 37 Class II YSOs within a $14' \times 10'$ region. 

Both clusters, Serpens South and W40, have been observed at $70-500~\mu$m 
with the \emph{Herschel} Space Telescope as part of the  Gould Belt program 
(Bontemps et al.\ 2010; Andr\'e et al.\ 2010; K{\"o}nyves et al.\ 2010).
The region observed toward the Aquila Rift is  $\sim 3.3^{\rm o}\times3.3^{\rm o}$ in size, 
with 7 YSOs safely classified as Class 0 objects, only in the reduced area of Serpens 
South (Bontemps et al.\ 2010). 
Additionally, around 45 (for $T_{\rm bol}^{70-500} < 27~\rm{K}$) and 60 
($L_{\rm submm}^{\lambda>350} /L_{\rm bol}^{70-500} > 0.03 $) objects in the entire 
field of Aquila (W40, Serpens South and the HII region
Sh2-62) were proposed to be Class 0 YSOs.
Maury et al.\ (2011) carried out a 1.2 mm dust continuum mapping 
of the Aquila complex with the MAMBO bolometer array on the IRAM 30m 
telescope. Twenty-five continuum sources were identified in the Serpens 
South protocluster and their evolutionary stages were estimated, resulting
in 9 starless sources, 9 Class 0  and 7 Class  0/I protostars.
In the W40 region, 35 sources were detected from these observations, 
separated into 14 starless, 8 Class 0, 4 Class 0/I and 9 Class I YSOs.

Additionally, a larger amount of YSOs have been identified in the Aquila/Serpens
region by the c2d (``from Molecular Cores to Planet-Forming Disks'') and 
GB (``\emph{Spitzer} Gould Belt Survey'') surveys (Dunham et al. 2013; Allen et al. 
in preparation). A total of 1524 YSOs with determined infrared spectral indexes 
and belonging to these regions are reported in their catalogs. 

Early estimates of the distance to the Serpens molecular cloud seemed to 
converge toward a value of 260 pc (see the discussion in Eiroa et al.\ 2008). 
On the basis of radial velocity measurements from
molecular line observations, Gutermuth et al.\ (2008) suggest that the Serpens
South cluster is comoving with the Serpens Main embedded cluster 
$3^{\circ}$ to the north, and, therefore, that it should be part of Serpens.
They also assign a distance of 260 pc to the cluster,
which corresponds to the distance to the
front edge of the Aquila Rift (Strai{\v z}ys et al.\ 1996).
Based on spectral types determined for a few main sequence stars in the 
W40 region, Shuping et al.\ (2012) estimated a distance  to the 
W40 cluster of between  455 and 535 pc.
Therefore, the W40 region and the Serpens South cluster (thought to be at 
260 pc) are usually regarded as separate objects. 
On the other hand, Bontemps et al.\ (2010)
argue that the W40 region, the Aquila Rift, and Serpens Main are parts of 
the same star-forming region, located at a common distance of 260 pc.
However, more recently, based on a comparison of the  X-ray Luminosity 
Function of the Serpens cluster with the previously published Orion Nebula cluster, 
Winston et al.\ (2010) obtained a  new distance for the Serpens core of
$360^{+22}_{-13}$ pc, while Dzib et al.\ (2010) claimed 
a distance of 415$\pm$5 pc to the same, based on a VLBA 
parallax of the embedded young AeBe star EC95. 
Given that the estimation of Dzib et al.\ (2010) is the most recent and 
accurate, in this paper we 
adopt a distance of 415 pc for the three regions in the Aquila complex, 
including Serpens, Serpens South, and W40.

The rest of this paper is organized as follows. In Section \ref{sec:obs} we present 
details of the  JVLA observations; in Section \ref{sec:results} we describe the results, which 
are  analyzed and discussed in Sections \ref{sec:discussion} and \ref{sec:comments}. 
Section \ref{sec:conclusions} is a summary of our results.

\section{Observations and data reduction}\label{sec:obs}

\begin{figure}[!ht]
\begin{center}
 \includegraphics[width=0.9\textwidth,angle=0]{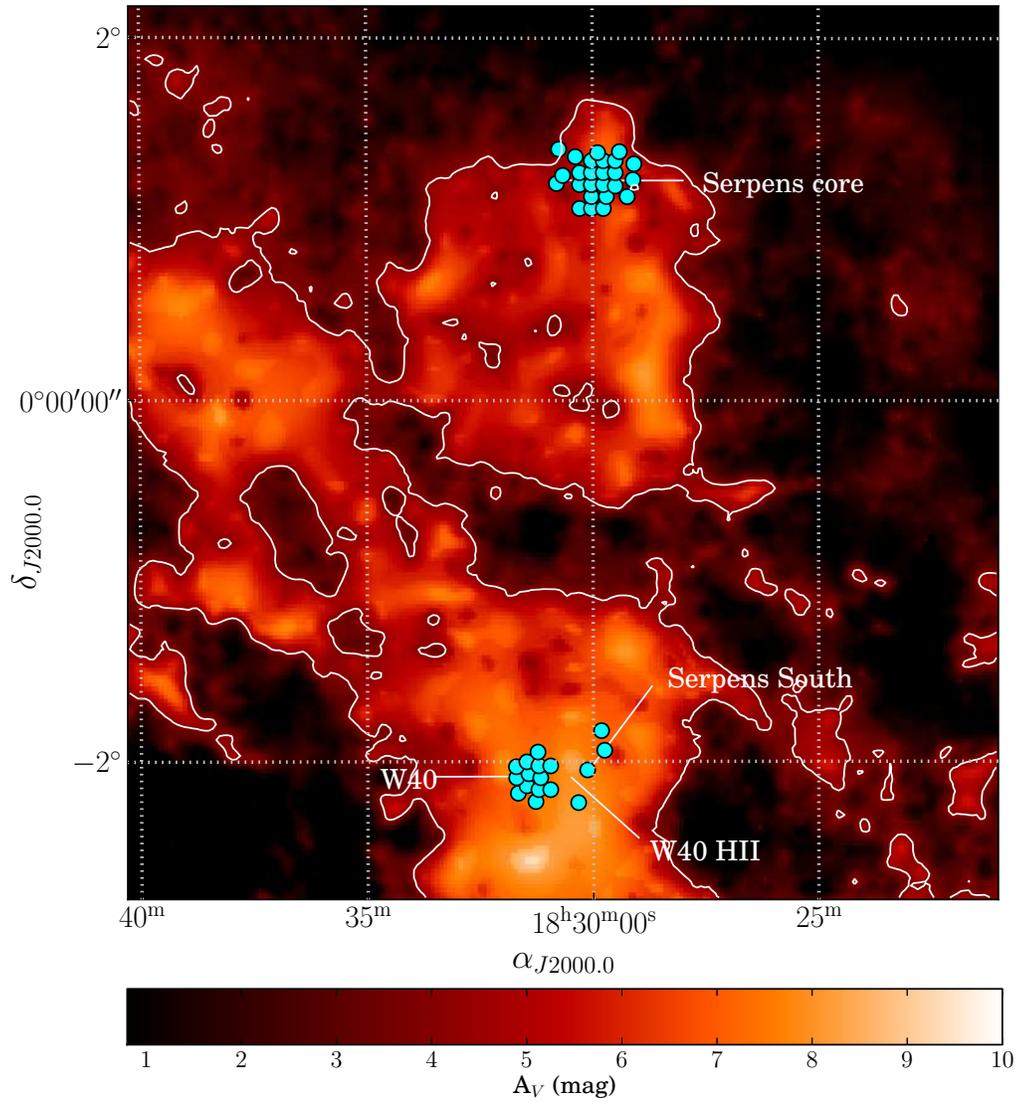}
\end{center}
 \caption{Extinction map of the Serpens star-forming region obtained
 as part of the COMPLETE project,  based on the 
STScI Digitized Sky Survey data (Cambr{\'e}sy 1999). The small cyan circles
represent the size  at 7.5 GHz of the fields observed in each region. }
\label{fig:map}
\end{figure}

The Serpens molecular cloud, the W40 region and the Serpens South cluster 
were observed with the JVLA in its A configuration. Two frequency 
sub-bands, each 1 GHz wide, and 
centered at 4.5 and 7.5 GHz, respectively, were recorded simultaneously.
The Serpens molecular cloud and the Serpens South cluster were observed in 
the same observing sessions on three different epochs (2011 June 17, July 19, and 
September 12 UT), using 25 and 4 pointings, respectively.
The  W40 region, on the other hand, was only observed on two epochs 
(2011 June 17 and July 16), using 13 pointings. 
This dual frequency strategy was chosen to enable the characterization 
of the spectral index of the detected sources, while the multi-epoch observations 
were aimed at determining the radio flux  variability and helping in the 
identification of the emission mechanisms (thermal vs. non-thermal). 
The details of the observations are listed in Table \ref{tab:obslog}.
The 25 pointings were used to map an area of 900 (530) square arcminutes at 4.5 
(7.5) GHz of the Serpens molecular cloud (see Figure \ref{fig:map}). 
The covered area  of the W40 region  using the 13 pointings was 415 
(280) square arcminutes at 4.5 (7.5) GHz.  The 4 pointings
used for the Serpens South cluster covered  an area of
290 (110) square arcminutes at 4.5 (7.5) GHz (see Figure \ref{fig:map}).
The number of individual pointings  observed toward
the Serpens molecular cloud and the W40 region, as well as the 
spacing between them, were chosen to optimize the compromise between 
uniform sensitivity and inclusion of the largest possible number of known 
young stars.

3C 286 and J1804+010 were used as the standard flux and phase calibrator,
respectively. The observations of the  Serpens Molecular cloud and the Serpens South cluster
were carried out in 2 hr scheduling blocks, while the observations of the W40 region
in, 1 hr scheduling blocks. Each epoch consists of a 9 minute observation of
the flux calibrator, followed by a series of two to four different target 
pointings (for 3 minutes each) bracketed by phase calibrator observations of 1 
minute. Thus, three minutes were spent on each target field in each epoch.
The data were edited and calibrated in a standard fashion 
using the Common Astronomy Software Applications package (CASA) version 3.4. 
Once calibrated, the data at each frequency were imaged (Stokes 
parameter $I$)  using the CASA task \texttt{clean}. 
The 25 target pointings toward the Serpens
Molecular cloud were used to construct a mosaic of 22400$\times$20412 pixels with 
a pixel size of $0.09''$ at 4.5 GHz. This was done by setting the  \emph{imagermode} parameter
to ``mosaic'' in the \texttt{clean} task. For the W40 region, the 13 target pointings 
were used to obtain a mosaic of 15552$\times$18522 pixels with a pixel size of $0.09''$
at 4.5 GHz. The phase centers of the mosaics are indicated in Table \ref{tab:obslog}. 
The 4 Serpens South cluster fields were imaged separately, using an image size of 
6750 pixels in each dimension and a pixel size of $0.09''$ at 4.5 GHz. The pixel size 
for the images at 7.5 GHz was $0\rlap.{''}055$, and the number of pixels
was adjusted to cover the same area as the images at 4.5 GHz. In order to take into 
account the non-coplanarity of the baselines far from the phase center, we set 
the \emph{gridmode} parameter to ``widefield'' with \emph{wprojplanes}=128 and \emph{facets}=1. 
We also correct the images for the primary beam attenuation. The rms noise levels 
reached at each frequency and epoch are given in Table \ref{tab:obslog}.

To produce images with improved sensitivity, the three or two epochs (in Serpens 
and W40 respectively) were combined  and jointly imaged. The rms noise levels
achieved after combining the epochs, as well as the synthesized beam (angular
resolution) of the final maps, are also given in Table \ref{tab:obslog}.

Gyrosynchrotron emission often (but not always) exhibits some level of circular 
polarization (Dulk 1985). To test for circular polarized radio emission,
we produced images of the  Stokes parameter $V$. The brightness distribution 
of the Stokes parameter V is obtained by applying the Fourier transform to the Stokes 
visibility  function $V_V$, which is given by $V_V = 1/2~ (V_{RR} - V_{LL})$, where 
$V_{RR}$ is the correlation of the right circularly polarized responses, 
and $V_{LL}$ is the correlation of those left circularly polarized.

We imaged the Stokes parameter $V$  for all the 42 fields (25 of the Serpens
molecular cloud, 4 of the Serpens South cluster, and 13 of the W40 region),
separately at each frequency, and combined the three or two epochs.
We did not  apply any polarization correction in 
addition to the standard 6 cm continuum calibration. 
Also, in order to avoid  beam squint  (Dzib et al.\ 2013), only the inner quarter 
(in area)  of the primary beam was tested, producing images 
of $4' \times 4'$ in size.

\section{Results}\label{sec:results}

\subsection{Source identification}\label{sec:si}

The identification of the radio sources was done through a visual inspection using 
the deep radio images. For the Serpens molecular cloud and the W40 region this means 
that we used the mosaics obtained after combining the epochs. For the Serpens South cluster
we used the combined individual fields. 

Once the sources were identified in those images, their fluxes as well as their 
positions at 4.5 and 7.5 GHz were obtained by performing 2-dimensional Gaussian fittings, 
using the CASA task \texttt{imfit}.  The results of these fittings are listed in 
Table \ref{tab:sources_sc}. 
Sources are named GBS-VLA J$hhmmss.ss-ddmmss.s$, where GBS-VLA stands for Gould's 
Belt Very Large Array Survey and $hhmmss.ss-ddmmss.s$ are the coordinates of the source. 
We considered for the quoted flux densities in the table three sources of error:
an error resulting from the statistical noise in the images, a systematic uncertainty 
of 5\% from possible errors in the absolute flux calibration, and the uncertainty induced by the  
pointing error of the VLA primary beam, which was included following Dzib et al.\ (2014).
Adding the three errors in quadrature, we obtained the total flux 
uncertainties listed in columns 3 and 5 of Table 2.
When a source is detected only at one frequency, an upper limit on the flux density of 
that source at the other band is given. The  upper limit corresponds to three times the 
rms noise of the area around the source position. We adopted the same criteria as 
Dzib et al.\ (2013) to consider a detection as firm. 
For new sources, i.e., those without reported counterparts in the literature,
we considered 5$\sigma$ detections, where $\sigma$ is the rms noise 
of the area around the source. For known sources with counterparts in the literature, 
on the other hand, we included 4$\sigma$ detections.
According to these criteria, we detected 94 sources in the Serpens 
molecular cloud, 41 in the W40 region, and 8 in the Serpens South cluster, for a total 
of 143 detections. Out of the 143 sources, 69 are new detections 
(see Section \ref{sec:counterparts}).

In order to compare the flux density between epochs and then to estimate the flux variability on 
a timescale of months, we searched in the images obtained from the individual epochs the 
sources detected in the combined ones. The flux densities and positions of the sources 
in these individual images are not shown for the sake of brevity. 

We investigated the number of false positives that could appear given the considerably 
large size of our mosaics. Assuming a Gaussian noise distribution, we estimated that
the number of possible false sources with a $5\sigma$ flux level are 7 at both 4.5 and 7.5 GHz
in the Serpens mosaic.  In the W40 mosaic, the number of possible false sources are 4 and 9 at 
4.5 and 7.5 GHz, respectively. For a $6\sigma$ flux level, we found that the number of false 
sources are below 0.03 for both mosaics and at both frequencies.
In Serpens, 5 sources   were detected  at 4.5 GHz with 
flux levels between $5-6\sigma$ and without counterparts at any other wavelengths (including 
counterparts at 7.5 GHz).  At 7.5 GHz, only one source without counterparts (including counterparts 
at 4.5 GHz) was detected with a $5-6\sigma$ flux level. Therefore, these 6 sources could be false 
detections.   In W40, all the sources without counterparts (including counterparts at 4.5 GHz or 
7.5 GHz) were detected above $6\sigma$, and therefore all sources reported in that region are 
real detections. We noted that in both mosaics, we considerer a radio source to have a counterpart 
at either 4.5 or 7.5 GHz, only if it is detected at  above $6\sigma$. 

\subsection{Source counterparts}\label{sec:counterparts}

GBS-VLA source positions were compared with source positions from X-ray, optical,
near-infrared, mid-infrared and radio catalogs. GBS-VLA sources were considered to have a 
counterpart at another wavelength when the positional coincidences were better than 
the combined uncertainties of the two datasets. These were about 1 arcsec for the 
infrared catalogs. For the X-ray and radio catalogs it depends on the instrument 
and its configuration.
The search was done in SIMBAD, and included all the major catalogs (listed explicitly 
in the footnote of  Table \ref{tab:sources_count1}).
We have also accessed the lists with all YSOs in the c2d-GB clouds compiled 
by Dunham et al.\ 2013 and Allen et al.\ (in preparation).
In total, 354 c2d-GB sources lie inside the regions observed by us. In order to 
find their radio counterparts, we imaged regions of 64 pixels in each dimension,
centered in the c2d-GB positions, and combining accordingly with each region, the three or 
two epochs. For this search we only used the field whose phase center was closest to the 
source. Three additional radio sources  were found in Serpens South in this pursuit, 
increasing the number of the radio detections to 146.

Out of 146 GBS-VLA sources, only 36 had previously been
detected at radio wavelengths (column 7 of Table \ref{tab:sources_count1}), while
the other 110 are new radio detections from this survey. On the other hand, we 
found a total of 63 counterparts at X-ray, near- and mid-infrared wavelengths, some of which 
have known radio counterparts.  In total, the number of sources that were previously known 
(at any frequency) is 77, while 69 of the sources in our sample are reported 
here for the first time. 

The classification of the sources shown in column 8 of Table \ref{tab:sources_count1} 
was taken from the literature, and is based on the IR and X-ray properties of the sources.
A total of 29 of the 77 sources with counterparts are firmly classified 
as YSOs, while HD 170634 (GBS-VLA J183024.87+011323.5) 
and W 40 IRS 5 (GBS-VLA J183114.82$-$020350.1) are typed as B7V and  
B1V stars, respectively. Two additional sources (GBS-VLA J183127.30$-$020504.5
and  J183127.45$-$020512.0) 
are considered  YSO candidates  (Rodr\'iguez et al.\ 2010). 
Three other sources have been associated with phenomena related to YSOs. 
Rodr\'iguez  et al.\ (2010) suggest that GBS-VLA J183128.67$-$020522.2 
could correspond to a shock front from a thermal jet (possibly powered by 
GBS-VLA J183128.65-020529.8) interacting with 
the ambient interstellar medium.  The sources GBS-VLA J183127.64$-$020513.5 and
J183127.67-020519.7 have been considered as ultracompact HII (UCHII) 
region candidates, centered around young massive stars. However, 
Shuping et al.\ (2012) argue that this  classification is unlikely, as the size of the 
unresolved radio sources at a distance of 500 pc would  be less than 100 AU, much 
smaller than a typical UCHII region (Kurtz 2005). Instead they propose that the  3.6 cm 
continuum flux could be due to free-free emission from shocked gas within 100 AU of 
the YSO caused by a jet or outflow.
We note  that, out of the 15 radio sources reported by Rodr\'iguez et al.\ (2010) with IR 
counterparts, seven were found by them not to be time variable (sources 
GBS-VLA J183114.82$-$020350.1, J183122.32$-$020619.6, J183127.64$-$020513.5,
J183127.67$-$020519.7, J183127.80$-$020521.9, J183128.01$-$020517.9,
and J183128.65$-$020529.8).
In our new observations, only GBS-VLA J183114.82$-$020350.1 and J183122.32$-$020619.6
are found to have high flux variability. However, when comparing our 7.5 GHz observations made in 
2011 to those made at 8.3 GHz in 2003 and 2004 by Rodriguez et al. (2010), remarkably, 
all sources show significant variations, typically by factors of $\sim$2 (see Appendix). 
We then conclude that the interpretation of 
the sources GBS-VLA J183127.64$-$020513.5 and J183127.67$-$020519.7
as steady UCHII regions (Rodriguez et al.\ 2010) 
is not correct, and that they are possibly gyrosynchrotron sources of slow time variability. 
If this is the case, they should be detectable as VLBI sources. 
The remaining 41 GBS-VLA sources with known counterparts at other wavelengths
are, to our knowledge, not classified in the literature.
In summary, we report a total of 110 unclassified sources, i.e., the 69 new detections 
plus the 41 sources that  have previously been detected  at other wavelengths but 
without a classification given in the literature.

\subsection{Spectral index, variability and circular polarization}\label{sec:vari}

An estimation of the radio spectral index $\alpha$ (the flux density $S_\nu \propto \nu^\alpha$)  
was obtained for most of the sources using 
$\alpha = \log (S_\nu\, (4.5) / S_\nu\, (7.5))/ \log (4.5~ \rm{GHz}/ 7.5~\rm{GHz} )$, 
where $S_\nu\,(4.5)$ and $S_\nu\, (7.5)$ are the flux densities at 4.5 and 7.5 GHz, 
respectively, and $\nu$ is the frequency of the incoming radiation. The spectral index is given in 
column 7 of Table \ref{tab:sources_sc}.

The repeated observations  allowed us to estimate the flux variation between 
the observed epochs at each frequency. Specifically, we determined the 
highest and lowest fluxes ($S_{\nu, \rm{max}}$ and $S_{\nu, \rm{min}}$, respectively) 
of the three or two epochs. The level of variability  was then 
estimated as the ratio of the difference between these values to the highest measured
flux, i.e., $\rm{variability } =(S_{\nu, \rm{max}}-S_{\nu, \rm{min}})/S_{ \nu, \rm{max}}$.
The resulting values, expressed in percentages, are given in columns 4 and 6 of Table 
\ref{tab:sources_sc}. The quoted uncertainties in the table for the
variability and spectral index were obtained, using standard error propagation theory, 
from the errors of the flux density (see Section \ref{sec:si}).
We considered as statistically significant only those variations that are above 
$3 \sigma$,  where $\sigma$ is the variability error of the source. In other words,
we consider a source as variable if the normalized difference between its highest 
and lowest flux density is greater than zero within an error of $3\sigma$.

Circular polarized radio emission was detected only from two sources in W40
(see Table \ref{tab:cir_pol}). The identification of these sources was done also
through a visual inspection in the Stokes $V$ images 
and searching around the position of the radio sources detected in 
Stokes $I$ emission. We required  the signal-to-noise ratio in the Stokes $V$ images 
to be greater than 5. Since the degree of circular polarization ($|V|/I$) of both 
sources is  lower than 10\%,  we cannot safely associate the
radio emission to gyrosynchrotron. 
As a consequence of the smaller size of the  Stokes $V$ images ($4' \times 4'$),
only 76 sources out of 143 were tested. The rest of them lie outside of the Stokes 
$V$ maps, but they could have circular polarization.

Given the existing deep X-ray, infrared, millimeter and sub-millimeter surveys carried out toward
these regions, it is unlikely that a large fraction of the 110 unclassified sources are
unidentified YSOs; however, we cannot rule out this possibility.
In order to characterize  the nature of the unclassified sources we analyze their radio 
properties. We find that 15 radio sources in the Serpens molecular cloud,
and  one source in the W40 region (Table \ref{tab:ysoC}) are compact, 
and have high levels of variability ($\apprge 50\%$ at a $3\sigma$ level) or a positive 
spectral index ($\geq +0.2$ within 1$\sigma$). 
Extragalactic sources do not usually
show high variability on such a short timescale (e.g. Hovatta et al.\ 2007, 2008; 
Lovell et al.\ 2008) 
or have a positive spectral index, while the radio emission seen in many Class III and some
youngest objects is highly variable (e.g. Feigelson 
\& Montmerle 1985; Feigelson et al.1998).

We propose then that these 16  unclassified sources are YSO candidates. 
The remaining 94 unclassified sources are then considered by us as extragalactic sources.
Notice that sources with spectral indices larger that -0.1 could also be due to free-free radiation 
(Rodr\'iguez et al.\ 1993). To firmly establish the nature of a radio continuum source we 
need observations of its morphology, spectral index, polarization, and time variability 
(Rodr\'iguez et al. 2012).

\section{Discussion}\label{sec:discussion}

\subsection{The lack of radio emission from Serpens South.}

Approximately 120 YSOs from the c2d-GB catalog lie inside the region mapped by us
in the Serpens South cluster. The detection of their radio counterparts is, surprisingly, 
very low. C2d-GB sources have only 3 radio counterparts in Serpens South (see 
Tables \ref{tab:sources_sc}  and \ref{tab:sources_count1}). These counterparts were 
detected with fluxes of $\apprle  0.10$ mJy (or radio luminosities of 
$\apprle 21\times10^{15}$ erg s$^{-1}$ Hz$^{-1}$, assuming a distance of 415 pc). Millimeter
dust continuum data in combination with  infrared Spitzer observations have suggested that 
the Serpens South cluster is very young (a few $10^5$ yr; Fern\'andez-L\'opez 
et al.\ 2014). It is then expected that the radio emission from most of the YSOs in the cluster 
is dominated by thermal emission from strong winds. Using equation (24) of Panagia \& Felli (1975) 
and following Rodr\'iguez et al. (1989), we estimated that for a spherical wind with a terminal velocity 
$v = 200 ~ \rm{km ~ s}^{-1}$, an electron temperature $T_e= 10^4$ K and a mass-loss rate of 
$10^{-7} ~ \rm{M}_\odot ~ \rm{yr}^{-1}$,  the radio flux density at a distance of 415 pc is 
$\sim 0.34$ mJy at 4.5 GHz. If the cluster is more distant, i.e. at 700 pc, the radio flux
decreases to $\sim 0.12$ mJy at 4.5 GHz, which is just equal to the sensitivity limit of our
observations toward that region ($5\sigma$ = $0.12$ mJy). It is then possible that we are not 
detecting the radio flux from the YSO population in Serpens South because the cluster is more 
distant than thought.

\subsection{Background sources}

We see from the previous section that a considerable fraction of the radio sources detected
are likely extragalactic objects (67 in the Serpens cloud, 19 in W40 and 8 in 
Serpens South, giving  a total of 94 sources).
In order to estimate the number of expected background radio sources, we follow
Anglada et al.\ (1998), who took into account the Gaussian primary-beam response of the 
VLA antennas. Considering that at 4.5 GHz the half-power width of the primary beam of the VLA
is  $\sim 10'$, we find that the number of expected radio sources in each
field with a flux greater than $S_{4.5}$ at 4.5 GHz  is given by
\begin{equation}\label{eq:back}
 N_{4.5}=1.21 \left(\frac{S_{4.5}}{\rm{mJy}}\right)^{-0.75}.
\end{equation}
%\noindent 
The 25 pointings used to map the Serpens molecular cloud correspond to a total area
of $25\cdot \pi \cdot (10'/2)^2 = 1963.5 $ square arcminutes at 4.5 GHz. However, because of the overlapping
between them, the effective covered area was 900  square arcminutes at 4.5 GHz, which is 
equivalent to 11 fields.

Therefore, using equation \ref{eq:back}, we find that $105$ background sources with a 
flux $\geq 64 ~\mu$Jy (5$\sigma$) at 4.5 GHz are predicted to lie 
within the mapped area of the Serpens molecular cloud.
The same exercise gives $40$ and $23$ background sources 
expected in the W40 region and the Serpens South cluster, respectively. Hence, we have detected
fewer extragalactic objects than expected by the count of Anglada et al. (1998) 
in the three regions, and we are statistically justified to assume that all of them  
are background and not associated with the region. Of course, this 
is subject to statistical variations.

\subsection{Radio properties of the YSO population}

\begin{figure}[!ht]
\begin{center}
 \includegraphics[width=0.7\textwidth,angle=0]{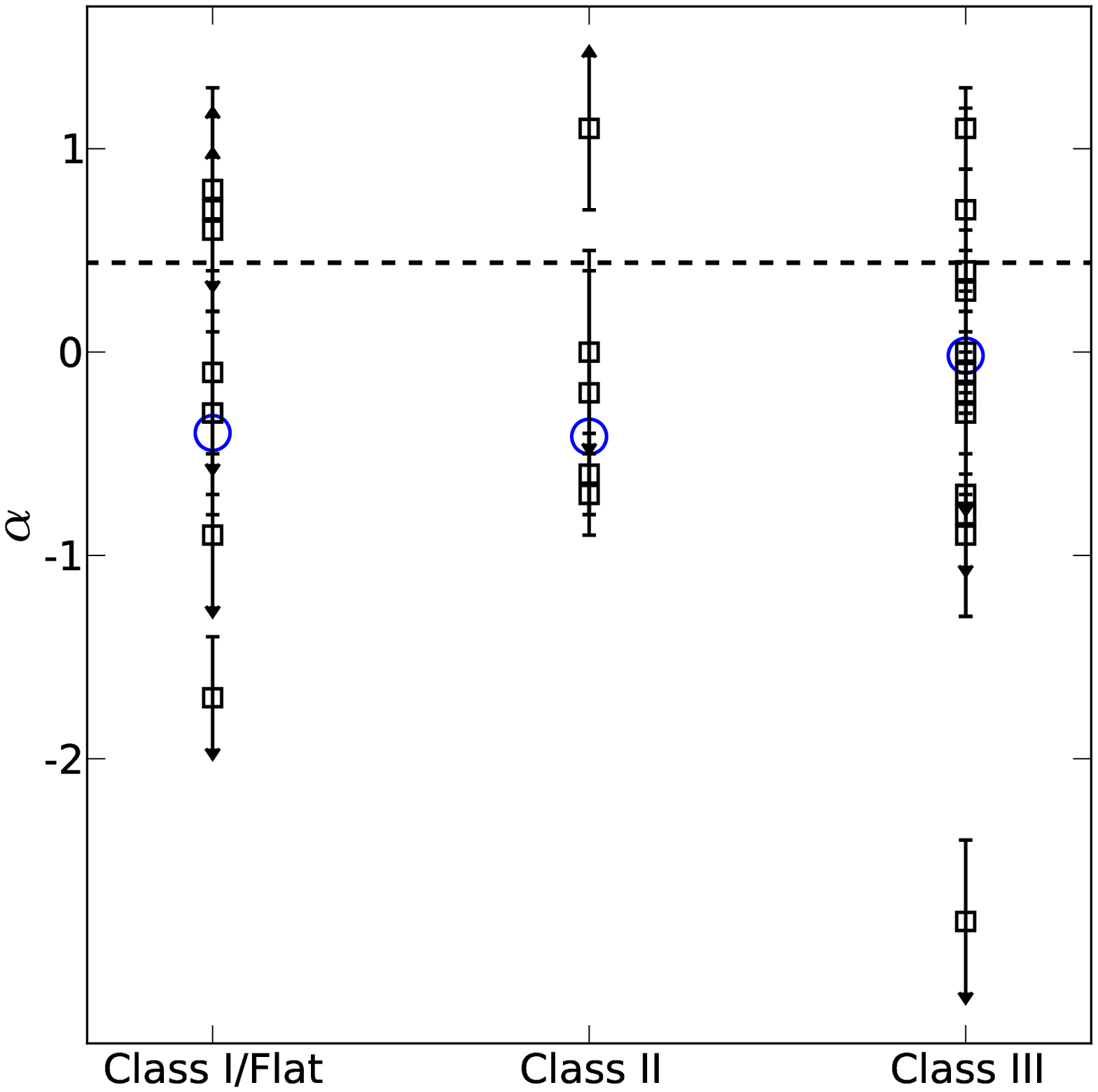}
\end{center}
 \caption{Spectral Index as a function of evolutionary status of the 25
 detected {Class I-III} YSOs. Values for the individual sources
 are shown with squares, and  weighted mean values with circles. The arrows 
 indicate upper and lower limits. The horizontal dashed line marks the weighted
 average spectral index of the YSO candidates listed in Table \ref{tab:ysoC}. }
\label{fig:spectral_index}
\end{figure}

We analyze the radio properties of the YSOs detected in our observations.
A subset of the detected YSOs with radio emission have a  SED classification reported
in the literature (see Table \ref{tab:yso} and references listed in its column 7).
Out of the 29 YSOs, 8 are Class I or Flat, 5 Class II and 12 Class III objects.
In Figure \ref{fig:spectral_index} we plot the spectral index of these 25 objects  
as a function of evolutionary status. 
The mean values of each  category are indicated by the large blue circles. 
We see that, given the large uncertainties involved,
it is not possible to distinguish between different evolutionary classes
based on the spectral index from flux densities at 4.5 and 7.5 GHz.
We note, however, that there are two Class I stars  with a very negative spectral 
index. This will be discussed later in section \ref{sec:comments}.
Notice also that most of the Class I YSOs we have detected have flux densities 
below $ 75~ \mu$Jy (6$\sigma$). For those sources, systematic errors are important,
and the  spectral index and variability errors are large.
For example, GBS-VLA J182957.60$+$011300.2 could have a  positive 
index ($+0.4$) if we consider 1$\sigma$ dispersion.

\begin{figure}[!ht]
\begin{center}
 \includegraphics[width=0.7\textwidth,angle=0]{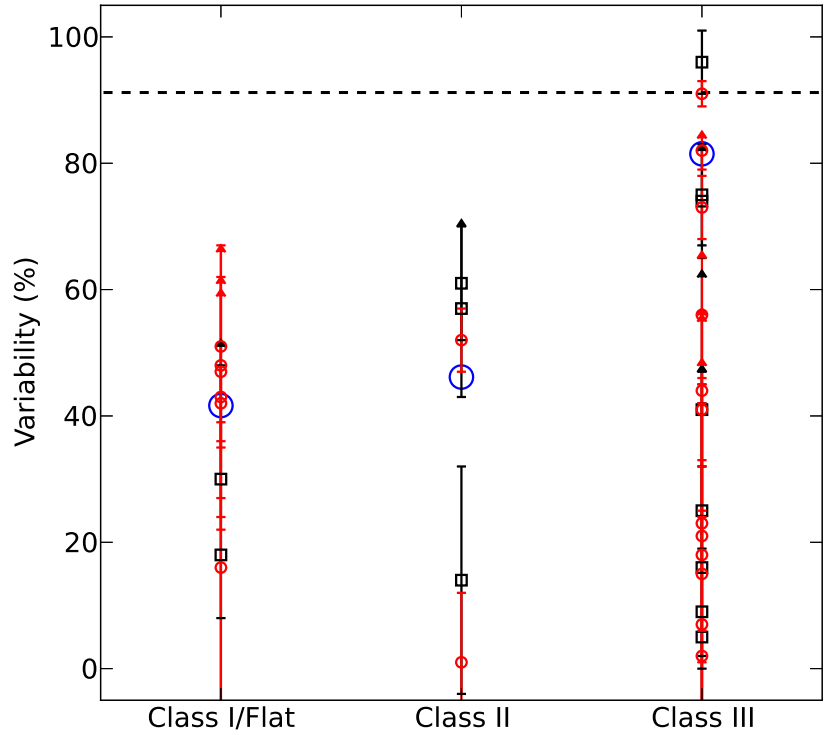}
\end{center}
 \caption{Variability  as a function of the evolutionary status 
 of 20 Class I-III YSOs. Values for the individual sources are shown 
 with red circles (variability at 4.5 GHz) and black squares (variability  at 7.5 GHz). 
 Some sources have a variability determination at both frequencies.
 The blue circles are the weighted average for each class. 
 The arrows indicate lower limits. The horizontal line marks the weighted average 
 variability at both frequencies of the YSO candidates listed in Table \ref{tab:ysoC}. }
\label{fig:variability}
\end{figure}

\begin{figure}[!ht]
\begin{center}
 \includegraphics[width=0.7\textwidth,angle=0]{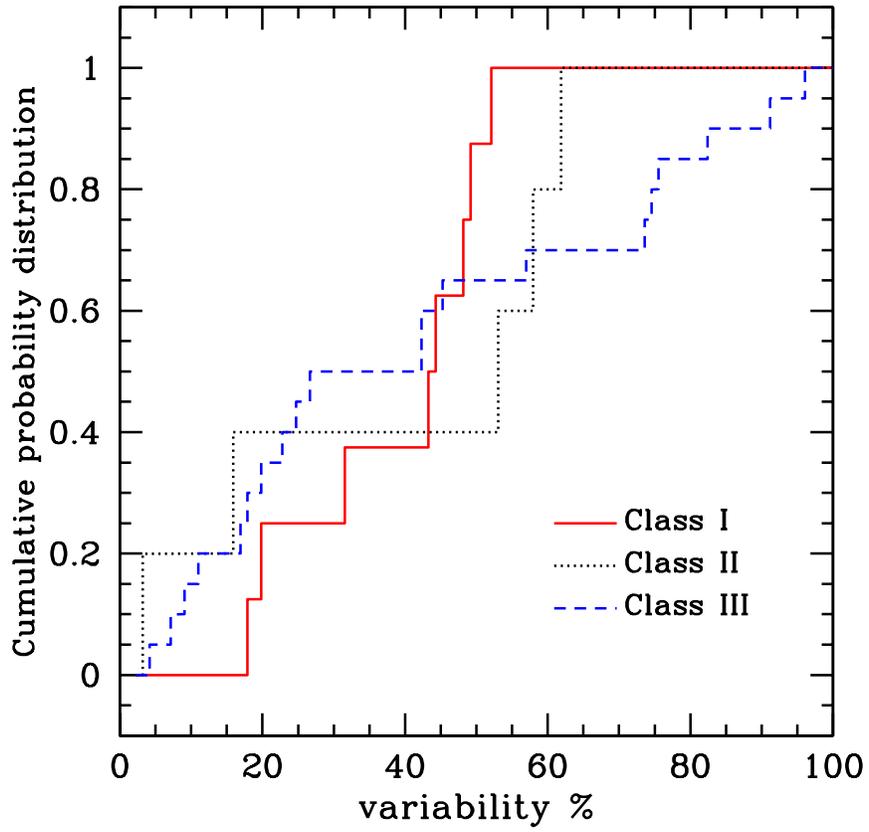}
\end{center}
 \caption{K-S test: cumulative probability distributions of variability in the
 three classes of YSOs. Red solid line: Class I; black dotted line: Class II;
 blue dashed line: Class III. }
\label{fig:ks}
\end{figure}

Out of the 25 detected Class I-III objects, 
20 have a variability determination at either 4.5 GHz or 7.5 GHz.
Figure \ref{fig:variability} shows the level of variability as 
a function of the evolutionary status for these 20 objects. 
We see that the older class (Class III) is populated by objects
with very high variable emission. In the same plot, we show
the weighted average of variability for each evolutionary class. While
Class I and Class II sources have weighted average variability lower than 50\%,
the weighted average variability of Class III is $\sim80\%$.
In order to test if variability increases with age, as is suggested by the plot, 
we carried out a  Kolmogorov-Smirnov (K-S) test
on the three YSO classes. The null hypothesis to test is that 
the variability distributions of different classes are drawn from
the same distribution. Figure \ref{fig:ks} shows the cumulative
probability distributions. The $D$ statistic gives the absolute maximum distance
between the cumulative distributions of two samples. The $D$ statistic
is $D_{\rm\footnotesize I-II}=1.16$ for class I and II distributions, 
$D_{\rm I-III}=0.89$ 
for class I and III distributions,  
and  $D_{\rm II-III} =0.65$ for class II and III distributions. 
The $p$-value gives the 
probability of obtaining the observed distributions when the null hypothesis
is true. We obtained a $p$-value of $p_{I-II} = 0.13$ for class I 
and II distributions,
$p_{I-III}=0.4$ for class I and III distributions, and $p_{II-III}=0.79$, 
for class II and III distributions. 
Assuming that sample pairs with $p<0.10$ are taken from different distribution
functions with high significance, we cannot reject the null hypothesis. This result
may be affected by the small number of YSOs in the samples, and the large uncertainties
of the variability. However we are interested in test this tendency in a forthcoming 
paper, using the whole sample of YSOs detected toward the Ophiuchus, Orion, 
Perseus and Taurus-Auriga regions.

We weighted average the variability of the YSO candidates identified in 
our observations (Table \ref{tab:ysoC}). 
This average is  shown as a horizontal line in Figure  \ref{fig:variability}. Interestingly, 
we find that the weighted average variability ($\sim  90\%$) of the YSO candidates is 
closest to the  averaged variability of the Class III known YSOs. 
This suggests that it might be a largest population of Class III sources out of 
our YSO candidates.

\subsection{The X-ray  -- radio relation}

For active stellar coronae, G{\"u}del \& Benz (1993) found a correlation 
between X-ray and radio luminosities, ${L_X}$
and ${L_R}$, respectively, which holds for X-ray luminosities over six 
orders of magnitude. The interpretation of this empirical relation is that the mechanism 
responsible for accelerating the non-thermal electrons that emit in the radio 
continuum also heats the coronal plasma, and this gives rise
to the thermal X-ray flux. Class Me dwarfs, Ke dwarfs and BY Dra stars, which typically have low 
luminosities, satisfy 
\begin{equation}\label{eq:gb1}
\frac{L_X}{L_R}  \approx 10^{15.5 \pm 1} ~~\rm{[Hz]}.
\end{equation}
\noindent More luminous classes (WTTS, RS CVn's binaries, Algols and FK Com stars) 
are systematically less X-ray bright compared to their radio luminosity, and fulfill
\begin{equation}\label{eq:gb2}
{L_X} / {L_R}  \lesssim  10^{15.5} ~~\rm{[Hz]}.
\end{equation}

We study the $L_X -L_R$ relation for the YSOs with X-ray counterparts in our sample. 
A total of 29 radio sources have  X-ray counterparts, 23 of which are YSOs.
However, we consider only YSOs with high radio variability or with a negative or flat spectral index 
(a subset of 18 Class I-III sources), thereby excluding sources that  could be not coronal. 
Also, we have corrected all X-ray luminosities to the distance of 415 pc adopted in this
work (see the discussion in Section \ref{sec:intro}). The corrected luminosity $L_0$
was obtained using $L_0 = (d_0/d)^2 L$, where $L$ is the 
luminosity of the source assuming a distance $d$, and $d_0$ 
the new adopted distance. 

\begin{figure}[!ht]
\begin{center}
 \includegraphics[width=0.7\textwidth,angle=0]{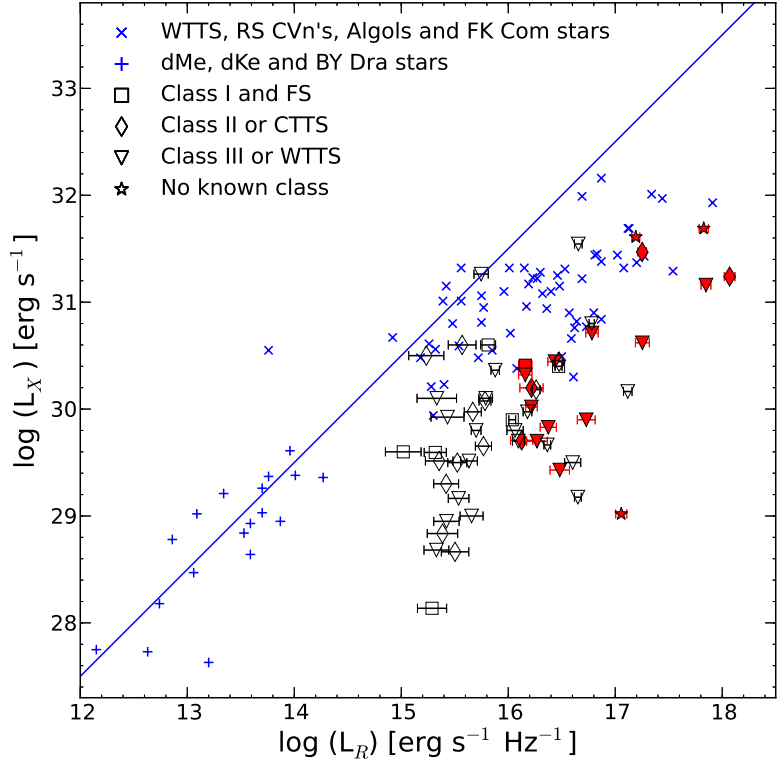}
\end{center}
 \caption{$L_X - L_R$ relation for stars  following G\"udel \& Benz (1993).
Symbols indicate different classes of stars as explained at the top left of the diagram.
Radio and X-ray luminosities of active stellar coronae represented by blue vertical
and diagonal crosses 
were taken from  G{\"u}del et al.\ (1993)  and Drake et al.\ (1989). 
The solid line with slope 1 is the fit obtained by G\"udel \& Benz (1993) 
for the dM(e), dK(e) and BY Dra stars (blue vertical crosses), which occupy 
the lower left portion of the diagram.
Open symbols correspond to YSOs detected 
in the Ophiuchus complex (Dzib et al.\ 2013). The YSOs detected by us
in the Serpens and W40 regions with X-ray
counterparts and with possible coronal radio emission are shown as red filled symbols.
The radio luminosity used for GBS-VLA sources is the average of the luminosities 
at 4.5 and 7.5 GHz. 
}
\label{fig:GB-relation}
\end{figure}

Following G{\"u}del \& Benz (1993), we place our subset of YSOs in the $L_X - L_R$ diagram  
(Figure \ref{fig:GB-relation}), and compare  with the relations already determined for stars of 
different classes  with magnetic activity.  Also plotted in Figure  \ref{fig:GB-relation} are 
the YSOs detected in the Ophiuchus complex by Dzib et al.\ (2013).
We find that the YSOs we have detected in Serpens and W40, as well as the sample of
YSOs in Ophiuchus, do not follow the  G{\"u}del-Benz relation for dwarf stars 
(equation \ref{eq:gb1}). Conversely, they fulfill $L_X / L_R   \lesssim 10^{15.5} ~\rm{[Hz]}$. 
G{\"u}del \& Benz (1993) proposed that the deviation from the relation with slope equal to
1 is likely a result of the sources having larger magnetospheres, which causes longer 
trapping times for the radio-emitting high-energy particles.

Lower X-ray fluxes than the  relation given by equation (\ref{eq:gb1}) could also be explained
if the photons are absorbed by  gas in front of the clouds. Kuhn et al.\ (2010) determined the visual 
absorption $A_{\rm V}$ toward their W40 sources from a \emph{J} vs. \emph{J}-\emph{H} diagram.
Our radio sources with counterparts in the catalog of Kuhn et al.\ (2010) have visual absorptions 
ranging from 6 to 22 magnitudes. Using the standard conversion 
 $N_H = (1.8\pm0.3)\times 10^{21} ~ {\rm cm}^{-2} \times A_{\rm V}$ (Predehl \& Schmitt 1995),
we find that the values of $A_{\rm V} = 6 - 22$ imply absorbing column densities of 
$N(\rm{H}) \sim 1-4\times10^{22}$ cm$^{-2}$. In the Serpens core, Giardino et al.\ (2007)
report column densities of $N(\rm{H}) \sim 0.5-7.4\times10^{22}$ cm$^{-2}$
for our YSOs with radio emission.
Another possibility for the deviation from relation (\ref{eq:gb1})
is a bias toward the brightest sources at the X-ray band in the G{\"u}del-Benz 
relation. According to this interpretation, the full area  below the relation could be populated 
with fainter coronal sources.
In fact, the YSOs in Serpens and Ophiuchus lie below the stars analyzed in the original study of 
G{\"u}del \& Benz (1993).

\subsection{Proper motions of YSOs in the W40 region}

\begin{figure}[!ht]
\includegraphics[width=1.1\textwidth,angle=0]{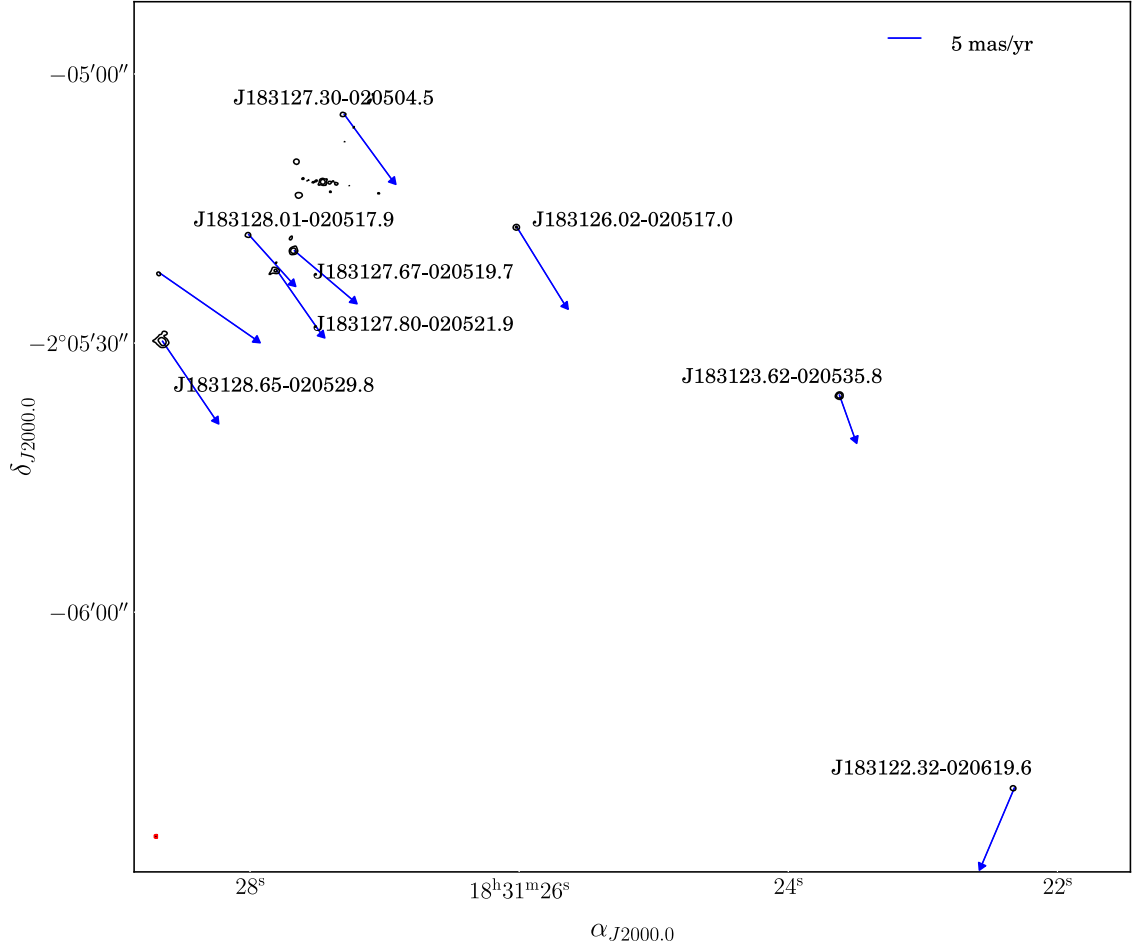}
\caption{6-cm radio continuum  images of the W40 region. The arrows indicate the
direction and length of the total proper motion of radio sources in the cluster,
detected by us and by Rodr\'iguez et al.\ (2010).
The $\theta_{\rm maj}  \times \theta_{\rm min} = 0\rlap.{''}41 \times
0\rlap.{''}40$, P.A. $= +75.8^{\rm o}$ synthesized beam is shown in the bottom
left of the map. Contours are 5, and 25 times 32 $\mu$Jy beam$^{-1}$, the rms noise 
of the image.}
\label{fig:w401}
\end{figure}

Rodr\'iguez et al.\ (2010) observed the W40 region at 3.6 cm with the VLA in its A and B array 
configurations. We use the observations obtained with the A configuration (2004 September 18)
to estimate the angular displacement of the sources between then and our own observations 
7 years later, in 2011.
The phase calibrator J1804$+$010 was used in all observing runs. 
In total, only 9 compact sources are detected at both epochs (see Table \ref{tab:prop_mov}). 
Notice that the displacement of the sources generated by their trigonometric parallax 
($p('')=1/D~[\rm{pc}]$) is 2.4 mas 
for a distance of $D=415$ pc, which is comparable to or even lower, by one order of magnitude, 
than the position errors of these sources ($2-15$ mas). Therefore, we do not consider 
the contribution from the parallax to the angular displacement of the sources.
The proper motions in right ascension, $\mu_\alpha \cos (\delta)$, and declination, $\mu_\delta$, are given 
in  Table \ref{tab:prop_mov} and plotted in Figure \ref{fig:w401}.
We also see that, with the exception of GBS-VLA J183122.32-020619.6 (RRR W40-VLA 3), all the 
sources are moving in the same direction with a mean absolute  value  $\mu_{total} =$ 
12.7 mas yr$^{-1}$. The average proper motions in right ascension and declination
of these 8 sources are $\mu_\alpha \cos (\delta) = -8.0$ mas yr$^{-1}$  and 
$\mu_\delta = -9.7 $ mas yr$^{-1}$. The sources GBS-VLA J183122.32-020619.6, J183123.62-020535.8,
J183126.02-020517.0, and J183128.65-020529.8 (RRR W40-VLA 3, 5, 8 and 18) are
associated with YSOs and, in fact, they all have non-zero proper motions.
Rodr\'iguez et al (2010) suggested that GBS-VLA J183127.30-020504.5 (RRR W40-VLA 9) is a 
YSO candidate, while we propose that GBS-VLA J183127.67-020519.7 (RRR W40-VLA 14) 
is a gyrosynchrotron source. The movements 
of these sources are similar to most of the stars  reported  in Table \ref{tab:prop_mov}, 
confirming that both objects are Galactic. It has been posited by Rodr\'iguez et al.\ (2010)
that  GBS-VLA J183128.67-020522.2 (RRR W40-VLA 19) is a shock front from a thermal 
jet (possibly powered by GBS-VLA J183128.65-020529.8). We note 
that this source has  not been detected at infrared, optical or X-rays wavelengths, which suggests that 
it is not a YSO. Its proper motion is the largest of the group and this could be due to 
an intrinsic motion.
The expected  proper motions for objects in the direction toward W40 at a distance of 415 pc 
are $\mu_\alpha \cos (\delta) = 0.76$ mas yr$^{-1}$  and  $\mu_\delta = -5.61 $ mas yr$^{-1}$.
These values are smaller than our estimations and probably this could be due to a 
systematic error (like an offset in the position of the phase calibrator in the archive data).
However, we have investigated this possibility and do not find any offset. It is also possible 
that the cluster has a peculiar velocity.

\section{Comments on individual sources}\label{sec:comments}

The source W 40 IRS 1d (GBS-VLA J183127.65$-$020509.7) was classified as a single source by Smith 
et al. (1985), but it has been recently resolved in the near-IR into a small cluster of 
at least 7 distinct sources by Shuping et al. (2012).
The position of the X-ray source associated with it, however, comes from a
high resolution ($0\rlap.{''}5$) X-ray image of the Chandra telescope.
The source is classified as an intermediate--mass YSO ($\sim 4~\rm{M}_{\odot}$;
Kuhn et al. 2010) and coincident, within the error, with the position of the 
radio source. Thus, we will consider that W 40 IRS 1d is associated to this 
young star. Also, this source does not show $K_s$-band-excess (Kuhn et al. 2010),
a fact that, along with its determined mass, suggests that the source may be an
HAeBe star.
The source W 40 IRS 5 (GBS-VLA J183114.82$-$020350.1) was first classified 
as a foreground star due to its lack of significant infrared absorption (Kuhn et al.\ 2010).
However, more recently, Shuping et al.\ (2012) classified this source as a B1V star. 
They also obtained an extinction toward it similar to that of other stars 
in the cluster. Moreover, the distance determined to this source ($\sim$469 pc)
is coincident with that determined to three additional OB stars in the region. This 
strongly  suggests that W 40 IRS 5  is part of the cluster and not a foreground star, as
proposed by Kuhn et al.\ (2010). 

\begin{figure}[!ht]
\begin{center}
 \includegraphics[width=0.9\textwidth,angle=0]{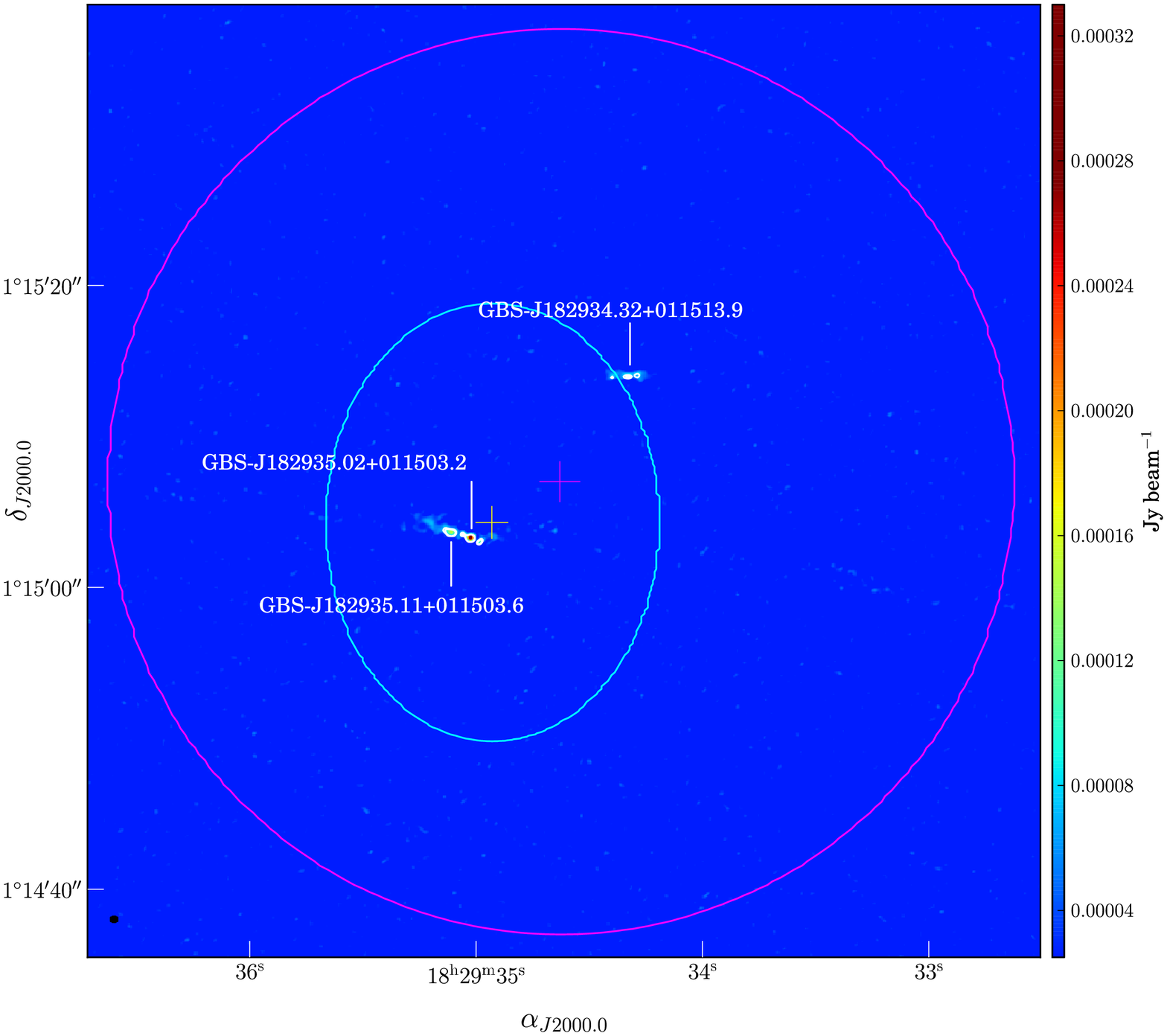}
\end{center}
 \caption{6-cm  radio continuum image of three GBS-VLA sources in the Serpens 
molecular cloud. The $\theta_{\rm maj}  \times \theta_{\rm min} = 0\rlap.{''}40 \times
0\rlap.{''}39$, P.A. $= +96.1^{\rm o}$ synthesized beam is shown in the bottom
left of the map. Contours are 5 and 6 times 15 $\mu$Jy beam$^{-1}$,
the rms noise of the image.
The cyan ellipse indicates the size of the source 
NVSS 182934$+$011504 ($29\rlap.{''}0 \times 22\rlap.{''}1$), from the
catalog of Condon et al.\ (1998).
The yellow cross marks, with error bars, the position of the NVSS source.
The  magenta circle indicates the size ($\sim 1'$) of the source 
\mbox{DCE08-210 5}, detected in the observations of Scaife et al. (2012) at 1.8 cm.
The magenta cross marks, with error bars, the position of the 1.8-cm source.
}
\label{fig:radio_c1}
\end{figure}

\begin{figure}[!ht]
\begin{center}
 \includegraphics[width=0.9\textwidth,angle=0]{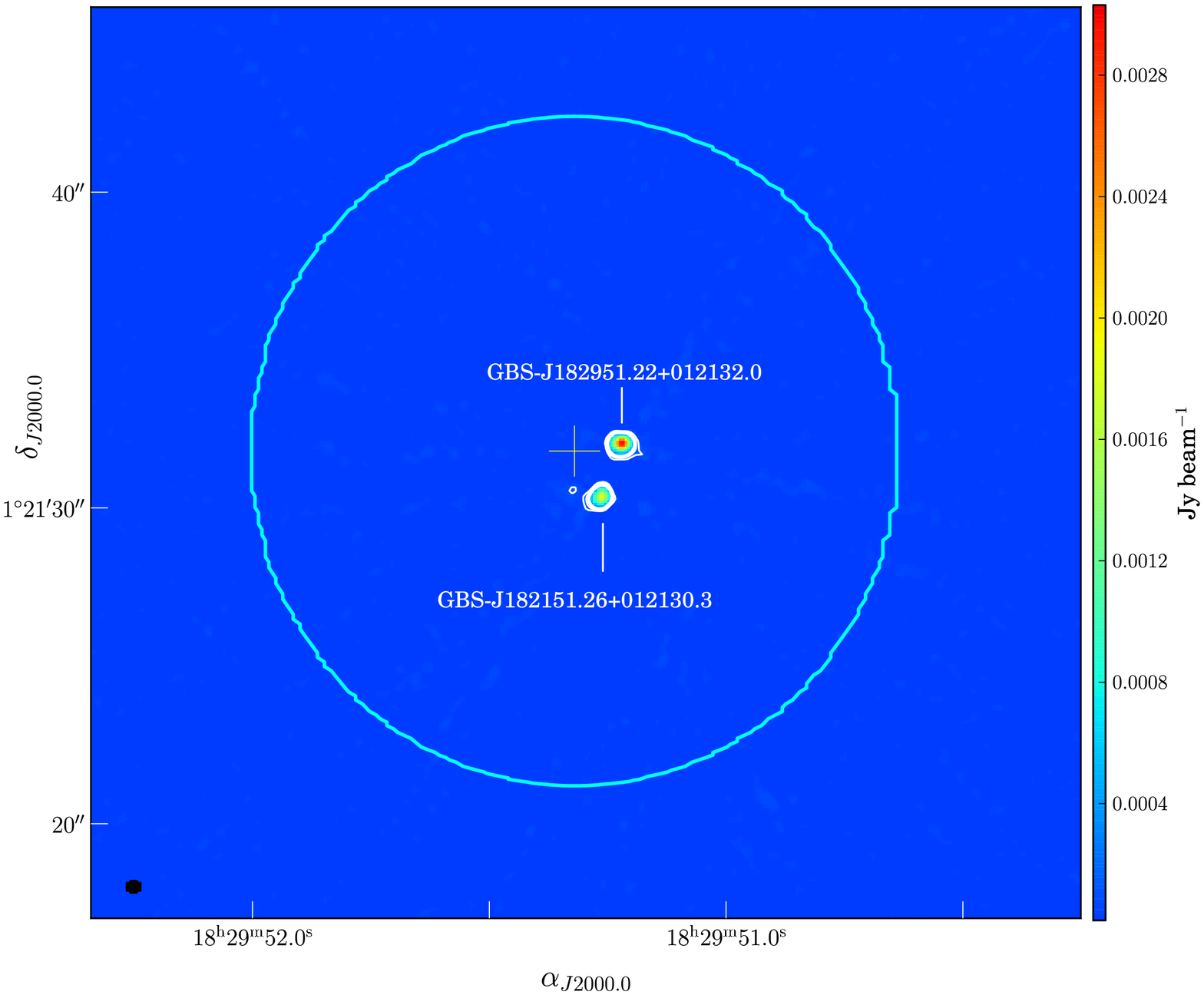}
\end{center}
 \caption{6-cm radio continuum image of two GBS-VLA sources in the Serpens 
molecular cloud. The $\theta_{\rm maj}  \times \theta_{\rm min} = 0\rlap.{''}40 \times
0\rlap.{''}39$, P.A. $= +96.1^{\rm o}$ synthesized beam is shown in the bottom
left of the map. Contours are 5, 10, 15 and 20 times 19 $\mu$Jy beam$^{-1}$,
the rms noise of the image.
The  cyan ellipse indicates the size of the source 
NVSS 182951$+$012131 ($21\rlap.{''}2 \times 20\rlap.{''}5$), from the
catalog of Condon et al.\ (1998).
The yellow cross marks, with error bars, the position of the NVSS source.
}
\label{fig:radio_c2}
\end{figure}

\begin{figure}[!ht]
\begin{center}
 \includegraphics[width=0.9\textwidth,angle=0]{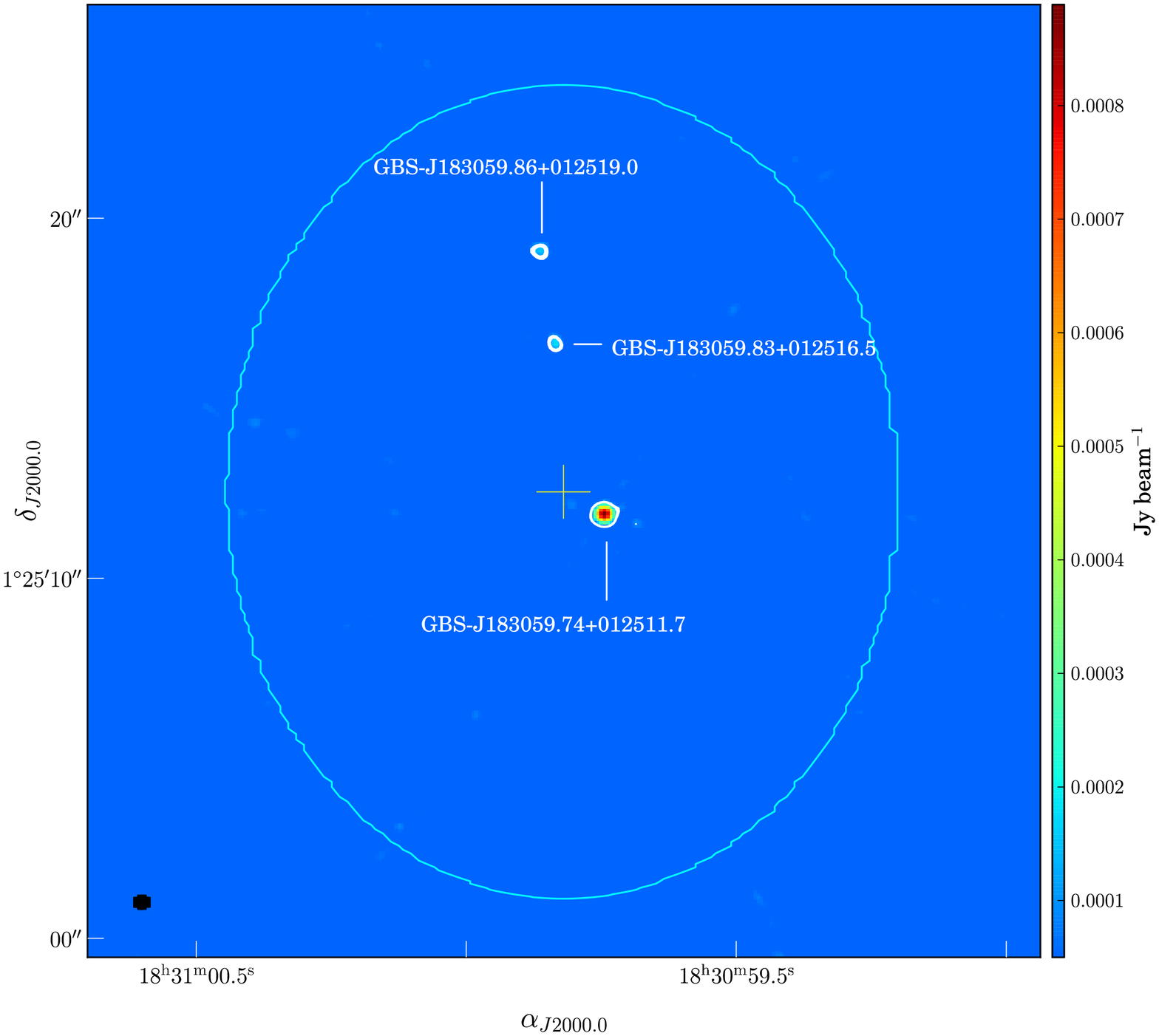}
\end{center}
 \caption{6-cm radio continuum  image of three GBS-VLA sources in the Serpens 
molecular cloud. The $\theta_{\rm maj}  \times \theta_{\rm min} = 0\rlap.{''}40 \times
0\rlap.{''}39$, P.A. $= +96.1^{\rm o}$ synthesized beam is shown in the bottom
left of the map. Contours are 5, 6 and 7 times 17 $\mu$Jy beam$^{-1}$,
the rms noise of the image.
The cyan  ellipse indicates the size of the source 
NVSS 183059$+$012512 ($22\rlap.{''}6 \times 18\rlap.{''}6$), from the
catalog of Condon et al.\ (1998).
The yellow cross marks, with error bars, the position of the NVSS source.
%NVSS 183059$+$012512.
}
\label{fig:radio_c3}
\end{figure}

\begin{figure}[!ht]
\begin{center}
 \includegraphics[width=0.9\textwidth,angle=0]{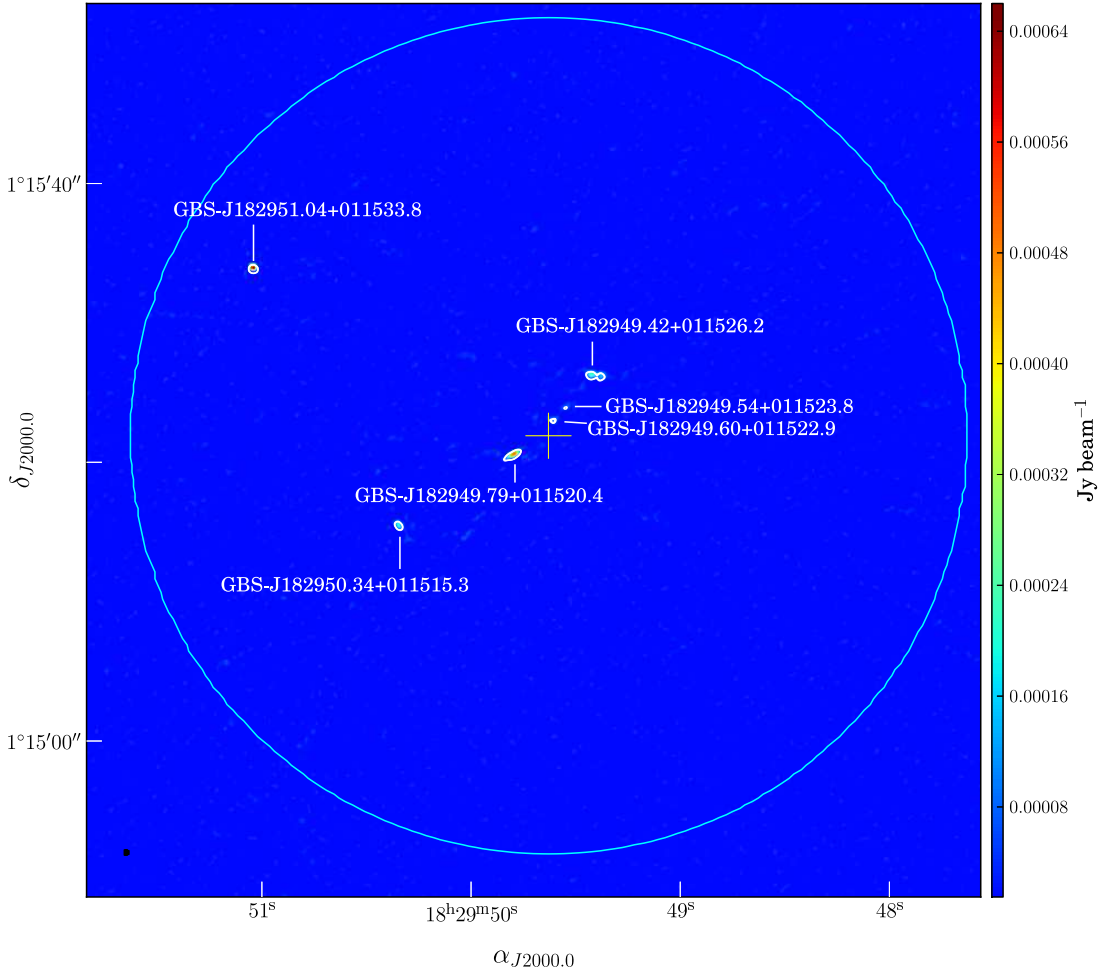}
\end{center}
 \caption{6-cm radio continuum  image of six GBS-VLA sources in the Serpens 
molecular cloud. The $\theta_{\rm maj}  \times \theta_{\rm min} = 0\rlap.{''}40 \times
0\rlap.{''}39$, P.A. $= +96.1^{\rm o}$ synthesized beam is shown in the bottom
left of the map. Contours are 5 and 6 times 16 $\mu$Jy beam$^{-1}$, the rms noise 
of the image. The  cyan circle indicates the size ($\sim 1'$) of the source 
DCE08-210 1, detected in the observations of Scaife et al. (2012) at 1.8 cm.
The yellow cross marks, with error bars, the position of the 1.8-cm source.
%DCE08-210 1.
}
\label{fig:radio_c4}
\end{figure}

The sources {NVSS 182934$+$011504} and {NVSS 182951$+$012131}, from the catalog of Condon et al.\ 
(1998), are resolved into double sources in our observations (Figures \ref{fig:radio_c1} 
and \ref{fig:radio_c2}), while NVSS 183059$+$012512 is resolved into a triple source 
(Figure \ref{fig:radio_c3}). Given the angular separation between the GBS-VLA sources 
and the peak of their NVSS counterparts as well as the uncertainties in their positions,
we associated {NVSS 182934$+$011504}, {NVSS 182951$+$012131}, and NVSS 183059$+$012512 with
GBS-VLA J182935.02$+$011503.2,  GBS-VLA J182951.22$+$012132.0, and GBS-VLA J183059.74$+$012511.7,
respectively (Table \ref{tab:sources_count1}). 
Three GBS-VLA sources lie inside the source DCE08-210 5 (size $\sim 1'$;
Figure \ref{fig:radio_c1}), detected by Scaife et al.\ (2012), and it is likely 
that this source includes multiple contributions, so it is difficult to attribute 
the 16 GHz emission conclusively to a specific GBS-VLA source. On the other hand, 
although DCE08-210 1 is resolved into 6 sources (Figure \ref{fig:radio_c4}), 
GBS-VLA J182949.79$+$011520.4 is the strongest source closer to the peak of the 16 GHz 
emission; therefore, we associate DCE08-210 5 with GBS-VLA J182949.79$+$011520.4.

\subsection{Non-thermal radio emission from YSOs}

In Table \ref{tab:yso} we list the radio properties  of the 29 YSOs detected. 
As we already mentioned, non-thermal gyrosynchrotron radio emission is characterized
by  high radio variability, and often  a negative spectral index  and some level of circular 
polarization. Out of the  12 Class III objects detected, 8 show high levels of variability 
or have a negative spectral index. We thus consider these 8 objects as possible sources 
of non-thermal emission.
Four Class II objects and, interestingly, four  Class I sources could also be
non-thermal. Additionally, the Herbig AeBe star GBS-VLA J183127.80$-$020521.9,
the proto-Herbig AeBe star (P-HAeBe) GBS-VLA J182957.89$+$011246.0, and the source 
GBS-VLA  J183128.01$-$020517.9 are likely  non-thermal radio sources. 
This gives us a high fraction ($65 \%$) of the YSOs detected in our observations
being non-thermal radio sources. An independent clue of the nature of the emission of these objects 
will be provided  by their detection (of lack thereof) in forthcoming VLBA observations.

The Class I objects with possible non-thermal emission
are GBS-VLA J182951.17$+$011640.4, J182952.22$+$011547.4, J182959.55$+$011158.1, and
J182959.94$+$011311.3.
Previously, some Class I objects in different star forming regions have been reported
as non-thermal emitters (Feigelson et al.\ 1998; Deller et al.\ 2013; Forbrich et al.\ 2007).
One explanation for the detection of non-thermal radio emission from these objects (which should be
absorbed by the ionized wind of the star)
is a geometrical effect.  According to this scenario, if the star is seen nearly pole-on or nearly edge-on, 
then the non-thermal radio emission originating in the corona might be less absorbed by the 
surrounding material and can reach the observer. Another possibility is
tidal clearing of circumstellar material in a tight binary system (Dzib et al.\ 2010).

\subsection{EC 95}

EC 95 = GBS-VLA J182957.89$+$011246.0 is a tight binary (angular separation of $\sim 15$ mas) 
consisting of a proto-Herbig AeBe star and a possibly low-mass T-Tauri companion. 
Both components were detected with the VLBA and  are therefore non-thermal radio sources
(Dzib et al.\ 2010). The origin of this non-thermal emission has been proposed to be intrinsic
magnetic activity in the stars. The magnetic activity in the low-mass T-Tauri companion is 
related to magnetic reconnection in the stellar surface. Electrons are then accelerated to 
mildly relativistic
velocities and generate gyrosynchrotron radiation. On the other hand, intermediate-mass stars are not 
expected to be magnetically active. Some processes have been suggested to explain the origin 
of the non-thermal radio emission in the proto-Herbig AeBe star (see Dzib et al.\ 2010 ), but still need to 
be tested. 
 
We derive a spectral index for the system of $0.1\pm0.2$, which is consistent 
within $1 \sigma$ with an early estimation by Smith et al.\ (2009), who found  $-0.26\pm0.26$.
We determined a high variability at 4.5 and 7.5 GHz ($ 65\, \%$ and $ 62\, \%$, 
respectively). We note that in the
VLBA observations obtained by Dzib et al.\ (2010) both components were found to be very variable,
at levels of  $\sim 94 \%$, and so the variability we derive should be associated with the intrinsic
variability of the stars. In conclusion, the radio properties we obtain for this system are
consistent with the non-thermal nature of the sources.

\section{Summary}\label{sec:conclusions}

We have carried out new  radio observations of three regions in the Aquila complex of local dark 
molecular clouds, namely the Serpens molecular cloud, the Serpens South cluster, and the W40 region.
We covered a large field of view ($\sim 0.45~{\rm deg}^2$) and, combined with high 
angular resolution ($\sim 0\rlap.{''}3$) and sensitivity ($\sim 17~\mu$Jy), our study
surpasses previous observations of these regions. We have detected a total of 146 sources. Twenty-nine
of them are associated with YSOs and 110 are new radio detections. The multi-epoch and dual frequency 
strategy allow us to speculate on the nature of the radio emission of the sources.  In particular, we find
that 16 of the  unclassified sources have a positive spectral index or exhibit high variability. They
might, hence, correspond to a small population of as-yet unidentified YSOs. What is more interesting is 
that approximately 65\% of the identified YSOs are non-thermal sources, of which 50\%  are bright enough
and therefore excellent targets for future astrometric observations with VLBI instruments.

\acknowledgments
This work is supported by CONACyT, Mexico, and PAPIIT, UNAM.
We thank M. Dunham and L. Allen for providing us with a list of known young stellar 
objects in the c2d-GB clouds prior to publication, and A. J. Maury for providing data
of the Serpens South cluster. The National Radio Astronomy Observatory is
operated by Associated Universities Inc. under cooperative agreement with the National
Science Foundation.

\appendix

\section{Appendix material}

For the 14 sources of the catalog by Rodr\'iguez et al.\ (2010) that are detected in our 
observations we carried out an independent analysis of variability. In order to compare the 
flux densities of our observations at 7.5 GHZ with the fluxes of Rodr\'iguez et al.\ (2010) 
at 8.3 GHz, we extrapolated using the corresponding spectral indices 
(see Table \ref{tab:rrr_sources}).
Unfortunately, in the paper of Rodr\'iguez et al.\ (2010) no errors are given for the flux 
densities of the sources (because the flux densities reported are the average of the two epochs 
observed). Assuming a typical error of 20\%, we find that the flux densities of three of the 
14 sources detected in both studies are consistent within $\pm 1\sigma$.
The remaining 11 sources show different flux densities for the two studies. Nine of the sources 
show a stronger flux density in the Rodr\'iguez et al.\ (2010) paper, while the remaining 2 show a 
stronger flux density in this paper. This diversity seems to rule out that the sources are steady 
and that there is a systematic calibration error in one of the two studies (because then we would 
expect all sources to appear as  brighter in one of the two studies). Finally, it should be stressed 
that in the Rodr\'iguez et al.\ (2010) paper, that reported observations in two different epochs 
separated by 0.88 yr (2003 November 3 and 2004 September 18), three of these 11 sources were 
already reported as time variable. Since the present paper offers a larger time baseline (about a decade), 
we expect the sources to exhibit even more variability between the two studies.

\begin{deluxetable}{lccccc}
\tabletypesize{\scriptsize}
\tablenum{A1}
\tablewidth{0pt}
\tablecolumns{6}
\tablecaption{W40 sources of the catalog of Rodr\'iguez
et al.\ (2010) detected in our observations.\label{tab:rrr}}
\tablehead{ GBS-VLA name & RRR W40-VLA  & RRR Flux density 
& GBS Flux density  & Spectral & RRR extrapolated flux density  \\
 & \colhead{number\tablenotemark{a}}& \colhead{[8.3 GHz]}& \colhead{[7.5 GHz] }  & 
\colhead{index}& \colhead{[7.5 GHz] } \\
& & \colhead{(mJy)} & \colhead{(mJy)} & & \colhead{(mJy)}\\
}
\startdata
%%%%%
J183114.82-020350.1 & 1 &  0.92$\pm$0.18   &     0.42$\pm$0.07 &  +0.3$\pm$0.2 &  0.89$\pm$0.18 \\
J183115.28-020415.2 & 2 &  0.90$\pm$0.18   &     0.44$\pm$0.07 &  -1.2$\pm$0.2  &  1.02$\pm$0.20\\
J183122.32-020619.6 & 3 &  0.47$\pm$0.09   &     1.13$\pm$0.22 &  +1.1$\pm$0.2  &  0.42$\pm$0.08\\
J183123.62-020535.8 & 5 &  3.97$\pm$0.79   &     3.32$\pm$0.45 &  -0.1$\pm$0.1  &  4.01$\pm$0.80\\
J183126.02-020517.0 & 8 &  3.27$\pm$0.65   &     0.75$\pm$0.09 &  -0.6$\pm$0.2  &  3.48$\pm$0.70\\
J183127.30-020504.5 & 9 &  0.99$\pm$0.20   &     0.45$\pm$0.07  & -0.2$\pm$0.3   &  1.01$\pm$0.20\\
J183127.45-020512.0 & 10 &  0.82$\pm$0.16  &      3.17$\pm$0.34 &  +0.3$\pm$0.2 &  0.80$\pm$0.16\\
J183127.64-020513.5 & 12 &  1.64$\pm$0.33  &      0.65$\pm$0.11 &  -0.6$\pm$0.3 &  1.74$\pm$0.35\\
J183127.65-020509.7 & 13 &  0.86$\pm$0.17  &      0.77$\pm$0.08 &  +0.1$\pm$0.2 &  0.85$\pm$0.17\\
J183127.67-020519.7 & 14 &  5.78$\pm$1.16  &      3.49$\pm$0.40 & +0.0$\pm$0.1  &  5.78$\pm$1.16\\
J183127.80-020521.9 & 15 &  1.71$\pm$0.34  &      1.24$\pm$0.15 &  -0.3$\pm$0.2 &   1.76$\pm$0.35\\
J183128.01-020517.9 & 16 &  0.94$\pm$0.19  &      0.50$\pm$0.07 &  -0.4$\pm$0.2 &   0.98$\pm$0.20\\
J183128.65-020529.8 & 18 &  11.1$\pm$2.22  &      4.79$\pm$0.71 &  -0.7$\pm$0.2 &   11.9$\pm$2.38\\
J183128.67-020522.2 & 19 &  0.20$\pm$0.04  &      0.25$\pm$0.04 &  -0.4$\pm$0.3 &   0.21$\pm$0.04\\
%\sidehead{b}
%%%%%
  \label{tab:rrr_sources}
\enddata
\tablenotetext{a}{The numbers in this column refer to the VLA source number in the catalog of Rodr\'iguez
et al.\ (2010).}
%\tablenotetext{b}{}
\end{deluxetable}

%%%%%%%%%%%%%%%%%%%%%%%%%%%%%%%%
%Tables and Figures:

\clearpage

\begin{deluxetable}{lcccllcc}
\tabletypesize{\scriptsize}
\rotate
\tablewidth{0pt}
\tablecolumns{8}
\tablecaption{JVLA observations \label{tab:obslog}}
\tablehead{ Region & Epoch\tablenotemark{1} &Time\tablenotemark{2} & Phase center & \multicolumn{2}{c}{Synthesized beam} 
& \multicolumn{2}{c}{rms noise\tablenotemark{3}} \\
\colhead{} & & \colhead{(UTC)}& \colhead{J2000}& \multicolumn{2}{c}{($\theta_{\rm maj}\times\theta_{\rm min}$; P.A.)}  & \multicolumn{2}{c}{($\mu$Jy beam$^{-1}$)} \\
\colhead{} & & \colhead{}& \colhead{}& \colhead{4.5 GHz}  & \colhead{7.5 GHz} & \colhead{4.5 GHz} & \colhead{7.5 GHz} \\
}
\startdata
{Serpens cloud (mosaic)} & 1 & Jun.17 07:59 & $18\rm{h}30\rm{m}00.0\rm{s} ~ +01\rm{d}12'37.0''$ & $0.42\times0.36$; $-$58.8 & $0.23\times0.22$; +100.7& 19.3 & 21.5\\
& 2 & Jul.19 04:25 &  & $0.49\times0.38$; +63.7 & $0.31\times0.23$; +66.9& 19.9 & 23.9\\
& 3 & Sep.12 03:47 &  & $0.39\times0.37$; +58.6 & $0.25\times0.23$; +50.4& 25.0 & 29.6\\
& C & --& & $0.40\times0.39$; +96.1 & $0.25\times0.23$; +62.2& 12.7 & 14.4 \\
W40 (mosaic)& 1 & Jun.17 09:30 & $18\rm{h}31\rm{m}20.0\rm{s} ~ -02\rm{d}05'00.0''$ & $0.45\times0.39$; +90.4 & $0.26\times0.23$; +54.2 & 22.8 & 28.0 \\
& 2 & Jul.16 06:35 & & $0.40\times0.39$; +12.0 & $0.25\times0.24$; +33.9 & 21.3 & 23.1\\
& C & --& & $0.41\times0.40$; +75.8 & $0.25\times0.23$; +40.5 & 16.0 & 18.3 \\
Serpens South (field 1) & 1 & Jun.17 07:59 & $18\rm{h}30\rm{m}19.9\rm{s} ~ -02\rm{d}13'42.0''$ & $0.46\times0.33$; +124.0 & $0.29\times0.21$; -56.6 & 38.2 & 29.9 \\
 & 2 & Jul.19 04:25 & & $0.53\times0.39$; +86.3 & $0.34\times0.24$; +86.8 &  41.5 & 32.4 \\
 & 3 & Sep.12 03:47 & & $0.37\times0.34$; -13.3 & $0.23\times0.22$; -40.0 & 47.8 & 33.8 \\
 & C & -- & & $0.44\times0.37$; -67.1 & $0.27\times0.23$; -70.3 & 24.0 & 18.5 \\
%%%
Serpens South (field 2) & 1 & Jun.17 07:59 & $18\rm{h}30\rm{m}07.8\rm{s} ~ -02\rm{d}02'49.1''$ & $0.45\times0.33$; -56.1 & $0.29\times0.21$; +123.4 & 39.1 & 31.0 \\
 & 2 & Jul.19 04:25 & & $0.53\times0.39$; +89.9 & $0.33\times0.24$; +85.3 & 41.9 & 34.3 \\
 & 3 & Sep.12 03:47 & & $0.37\times0.34$; -13.0 & $0.23\times0.22$; -36.9 & 48.1 & 36.5 \\
 & C & -- & & $0.44\times0.37$; +114.2 & $0.27\times0.23$; -70.7 & 24.6 & 19.4 \\
%%%
Serpens South (field 3) & 1 & Jun.17 07:59 & $18\rm{h}29\rm{m}49.2\rm{s} ~ -01\rm{d}49'45.0''$ & $0.45\times0.33$; -57.5 & $0.23\times0.22$; -50.7 & 38.4 & 28.7 \\
 & 2 & Jul.19 04:25 & & $0.52\times0.39$; +83.0 & $0.33\times0.24$; +83.1 & 37.8 & 31.8 \\
 & 3 & Sep.12 03:47 & & $0.35\times0.35$; -36.5 & $0.23\times0.22$; -52.0 & 47.1 & 33.6 \\
 & C & -- & & $0.43\times0.37$; -72.0 & $0.25\times0.23$; +94.2 & 23.3 & 18.1 \\
%%%
Serpens South (field 4) & 1 & Jun.17 07:59 & $18\rm{h}29\rm{m}45.1\rm{s} ~ -01\rm{d}56'17.4''$ & $0.45\times0.33$; +123.5 & $0.23\times0.22$; -45.9 & 39.6 & 29.3 \\
 & 2 & Jul.19 04:25 & & $0.51\times0.39$; +80.4 & $0.33\times0.24$; +82.0 & 41.0 & 31.5 \\
 & 3 & Sep.12 03:47 & & $0.37\times0.34$; +2.1& $0.23\times0.22$; -25.0 & 74.1 & 73.0 \\
 & C & -- & & $0.45\times0.38$; -72.2 & $0.26\times0.23$; +93.9 & 27.1 & 20.9 \\
\enddata
\tablenotetext{1}{C indicates parameters measured in the images after combining the epochs.}
\tablenotetext{2}{Start time of the observing sessions. All epochs were observed in 2011.}
\tablenotetext{3}{Measured at the center of the Stokes $I$ image.}
\end{deluxetable}

\begin{deluxetable}{lcrrrrr}
%\rotate
\tabletypesize{\scriptsize}
\tablewidth{0pt}
\tablecolumns{7}
\tablecaption{Radio sources detected in Serpens }
\tablehead{GBS-VLA & New  & Flux density &Variability & Flux density & Variability&Spectral\\
\colhead{ Name/Position} & detection?\tablenotemark{1} & \colhead{ [4.5 GHz]}
& \colhead{ [4.5 GHz]} & \colhead{[7.5 GHz]} & \colhead{[7.5 GHz] } & \colhead{Index}\\
& &\colhead{(mJy)} &  \colhead{(\%)} & \colhead{(mJy)} & \colhead{(\%)} & \\
}
%\colhead{\multicolumn{6}{c}{Detected sources in the Serpens molecular cloud} }\\
\startdata
\sidehead{Detected sources in the Serpens molecular cloud:}
J182850.71+011102.7 & Y & 0.28 $\pm$ 0.05 &  38 $\pm$  19 & -- & -- & -- \\
J182851.30+010908.6 & N & 8.77 $\pm$ 1.72 &  14 $\pm$  24 & -- & -- & -- \\
J182851.48+010947.3 & Y & 0.16 $\pm$ 0.03 &  50 $\pm$  17 & -- & -- & -- \\
J182854.44+011859.7 & Y & 0.18 $\pm$ 0.02 &   6 $\pm$  22 & 0.12 $\pm$ 0.03 &  25 $\pm$  31 & -0.9 $\pm$  0.3 \\
J182854.46+011823.7 & Y & 3.13 $\pm$ 0.32 &  44 $\pm$   9 & 5.72 $\pm$ 1.18 &  17 $\pm$  24 &  1.2 $\pm$  0.1 \\
J182854.87+011753.0 & Y & 0.08 $\pm$ 0.02 & $>$44 $\pm$  19 & 0.10 $\pm$ 0.03 & --A-- &  0.4 $\pm$  0.5 \\
J182901.40+010434.7 & N & 0.45 $\pm$ 0.08 &  26 $\pm$  19 & -- & -- & -- \\
J182903.06+012331.0 & Y & 0.36 $\pm$ 0.07 &  25 $\pm$  21 & -- & -- & -- \\
J182903.13+010346.0 & Y & 0.06 $\pm$ 0.02 & --A-- & -- & -- & -- \\
J182905.07+012309.0 & N & 0.12 $\pm$ 0.02 &  38 $\pm$  20 & -- & -- & -- \\
J182906.84+011742.7 & N & 6.46 $\pm$ 0.39 &   8 $\pm$   8 & 6.10 $\pm$ 0.53 &  14 $\pm$  11 & -0.1 $\pm$  0.1 \\
J182907.07+011801.9 & Y & 2.82 $\pm$ 0.18 & Extended & 1.67 $\pm$ 0.16 & Extended & -1.1 $\pm$  0.2 \\
J182907.62+012125.1 & N & 0.10 $\pm$ 0.02 &  29 $\pm$  22 & $<$0.04 & -- & $<$-1.9 $\pm$  0.3 \\
J182910.17+012559.5 & N & 1.41 $\pm$ 0.25 &  71 $\pm$   8 & -- & -- & -- \\
J182911.94+012119.4 & Y & 0.35 $\pm$ 0.05 &  28 $\pm$  16 & 0.41 $\pm$ 0.11 &  18 $\pm$  31 &  0.3 $\pm$  0.2 \\
J182912.01+011415.1 & Y & 0.08 $\pm$ 0.01 & $>$14 $\pm$  21 & 0.07 $\pm$ 0.02 & $>$45 $\pm$  25 & -0.2 $\pm$  0.5 \\
J182913.17+010906.4 & N & 0.51 $\pm$ 0.04 &  19 $\pm$  12 & 0.60 $\pm$ 0.08 &   8 $\pm$  18 &  0.3 $\pm$  0.2 \\
J182913.36+011544.3 & Y & 0.20 $\pm$ 0.03 &  19 $\pm$  20 & 0.07 $\pm$ 0.02 & $>$40 $\pm$  25 & -2.1 $\pm$  0.5 \\
J182913.79+010738.6 & N & 0.06 $\pm$ 0.01 &  36 $\pm$  25 & 0.10 $\pm$ 0.02 & $>$26 $\pm$  20 &  0.9 $\pm$  0.5 \\
J182916.11+010437.5 & N & 0.21 $\pm$ 0.03 &  41 $\pm$  16 & 0.25 $\pm$ 0.06 &  52 $\pm$  19 &  0.4 $\pm$  0.2 \\
J182918.23+011757.7 & N & 0.38 $\pm$ 0.05 &  32 $\pm$  15 & 0.26 $\pm$ 0.07 &  41 $\pm$  24 & -0.8 $\pm$  0.2 \\
J182926.71+012342.1 & N & 0.18 $\pm$ 0.02 &  30 $\pm$  13 & 0.11 $\pm$ 0.02 &  34 $\pm$  25 & -1.0 $\pm$  0.4 \\
J182928.02+011156.5 & N & 0.19 $\pm$ 0.02 &  18 $\pm$  18 & 0.09 $\pm$ 0.02 & --A-- & -1.5 $\pm$  0.5 \\
J182928.28+011205.7 & N & 0.21 $\pm$ 0.04 &  52 $\pm$  19 & $<$0.05 & -- & $<$-3.1 $\pm$  0.4 \\
J182929.78+012158.1 & N & 0.07 $\pm$ 0.01 & --A-- & 0.06 $\pm$ 0.02 & --A-- & -0.2 $\pm$  0.6 \\
J182930.71+010048.3 & N & 32.10 $\pm$ 5.60 & Extended & -- & -- & -- \\
J182932.21+012104.6 & Y & 0.10 $\pm$ 0.01 &  52 $\pm$  15 & 0.06 $\pm$ 0.02 & $>$20 $\pm$  27 & -0.8 $\pm$  0.5 \\
J182933.07+011716.3 & N & 0.27 $\pm$ 0.03 &  73 $\pm$   5 & 0.32 $\pm$ 0.05 &  74 $\pm$   9 &  0.3 $\pm$  0.2 \\
J182934.12+010810.9 & Y & 2.06 $\pm$ 0.17 &  16 $\pm$  11 & 2.14 $\pm$ 0.33 &  20 $\pm$  18 &  0.1 $\pm$  0.1 \\
J182934.32+011513.9 & N & 0.73 $\pm$ 0.13 & Extended & 0.12 $\pm$ 0.02 & $>$48 $\pm$  12 & -3.7 $\pm$  0.5 \\
J182935.02+011503.2 & N & 0.35 $\pm$ 0.05 &  21 $\pm$  15 & 0.21 $\pm$ 0.03 &  34 $\pm$  18 & -1.0 $\pm$  0.3 \\
J182935.11+011503.6 & N & 0.50 $\pm$ 0.07 & Extended & $<$0.05 & -- & $<$-4.8 $\pm$  0.3 \\
J182936.50+012317.0 & N & 0.29 $\pm$ 0.04 &  35 $\pm$  14 & 0.21 $\pm$ 0.05 &  39 $\pm$  24 & -0.7 $\pm$  0.2 \\
J182937.76+010314.6 & N & 0.60 $\pm$ 0.06 &  13 $\pm$  16 & 0.33 $\pm$ 0.06 &  32 $\pm$  19 & -1.2 $\pm$  0.2 \\
J182938.87+011850.4 & Y & 0.15 $\pm$ 0.02 &  30 $\pm$  16 & 0.13 $\pm$ 0.02 &  35 $\pm$  23 & -0.4 $\pm$  0.3 \\
J182939.09+011233.6 & Y & 0.13 $\pm$ 0.02 &   9 $\pm$  24 & 0.05 $\pm$ 0.02 & $>$19 $\pm$  33 & -1.9 $\pm$  0.6 \\
J182940.03+011051.2 & Y & 0.09 $\pm$ 0.01 & $>$32 $\pm$  15 & 0.09 $\pm$ 0.02 &  45 $\pm$  23 &  0.0 $\pm$  0.4 \\
J182944.07+011921.1 & N & 4.49 $\pm$ 0.24 &  21 $\pm$   6 & 4.64 $\pm$ 0.27 &   5 $\pm$   8 &  0.1 $\pm$  0.1 \\
J182948.83+010647.4 & N & 0.50 $\pm$ 0.05 &  29 $\pm$  13 & 0.76 $\pm$ 0.13 &   6 $\pm$  23 &  0.8 $\pm$  0.2 \\
J182948.92+011523.8* & Y & 0.06 $\pm$ 0.01 & $>$62 $\pm$  10 & $<$0.04 & -- & $<$-0.5 $\pm$  0.5 \\
J182949.42+011526.2 & Y & 0.72 $\pm$ 0.09 & Extended & 0.28 $\pm$ 0.05 & Extended & -1.9 $\pm$  0.4 \\
J182949.50+011955.8 & Y & 2.44 $\pm$ 0.16 &  18 $\pm$   8 & 1.63 $\pm$ 0.16 &   7 $\pm$  13 & -0.8 $\pm$  0.1 \\
J182949.54+011523.8 & Y & 0.16 $\pm$ 0.02 &  76 $\pm$   7 & $<$0.04 & -- & $<$-2.5 $\pm$  0.3 \\
J182949.60+011522.9 & Y & 0.18 $\pm$ 0.02 &  71 $\pm$   8 & 0.14 $\pm$ 0.03 &  64 $\pm$  14 & -0.5 $\pm$  0.4 \\
J182949.79+011520.4 & N & 0.88 $\pm$ 0.07 &  43 $\pm$   9 & 0.99 $\pm$ 0.10 &  23 $\pm$  13 &  0.2 $\pm$  0.2 \\
J182950.34+011515.3 & Y & 0.33 $\pm$ 0.04 &  11 $\pm$  20 & 0.20 $\pm$ 0.04 & $>$61 $\pm$  18 & -1.0 $\pm$  0.4 \\
J182951.04+011533.8 & N & 0.61 $\pm$ 0.05 &  41 $\pm$   9 & 0.58 $\pm$ 0.07 &  46 $\pm$  11 & -0.1 $\pm$  0.2 \\
J182951.17+011640.4 & N & 0.09 $\pm$ 0.02 & $>$51 $\pm$  16 & 0.07 $\pm$ 0.02 & $>$30 $\pm$  22 & -0.3 $\pm$  0.5 \\
J182951.17+010529.7 & Y & 0.09 $\pm$ 0.01 & --A-- & $<$0.04 & -- & $<$-1.7 $\pm$  0.3 \\
J182951.22+012132.0 & N & 3.80 $\pm$ 0.22 &  20 $\pm$   7 & 2.09 $\pm$ 0.17 &   6 $\pm$  11 & -1.2 $\pm$  0.2 \\
J182951.26+012130.3 & Y & 2.46 $\pm$ 0.16 &  26 $\pm$   8 & 1.56 $\pm$ 0.13 &   9 $\pm$  12 & -0.9 $\pm$  0.2 \\
J182952.22+011547.4 & N & 0.12 $\pm$ 0.02 &  16 $\pm$  23 & $<$0.05 & -- & $<$-1.7 $\pm$  0.3 \\
J182953.99+011229.5 & N & 0.08 $\pm$ 0.01 &  46 $\pm$  19 & $<$0.04 & -- & $<$-1.4 $\pm$  0.3 \\
J182954.30+012011.2 & Y & 0.14 $\pm$ 0.02 & $>$75 $\pm$   5 & $<$0.04 & -- & $<$-2.3 $\pm$  0.3 \\
J182954.31+010309.6 & N & 0.59 $\pm$ 0.05 &  17 $\pm$  14 & 0.60 $\pm$ 0.09 &  11 $\pm$  20 &  0.0 $\pm$  0.2 \\
J182954.36+010350.4 & Y & 0.07 $\pm$ 0.01 & $>$22 $\pm$  18 & $<$0.04 & -- & $<$-0.8 $\pm$  0.4 \\
J182954.86+011129.3 & Y & 0.10 $\pm$ 0.01 &  61 $\pm$  21 & 0.07 $\pm$ 0.02 & $>$22 $\pm$  26 & -0.7 $\pm$  0.5 \\
J182955.76+010440.3 & Y & 0.14 $\pm$ 0.02 &  14 $\pm$  21 & 0.12 $\pm$ 0.02 &  28 $\pm$  26 & -0.4 $\pm$  0.3 \\
J182956.96+011247.6 & N & $<$0.04 & -- & 0.08 $\pm$ 0.02 & $>$57 $\pm$  14 & $>$ 1.1 $\pm$  0.4 \\
J182957.60+011300.2 & N & 0.05 $\pm$ 0.01 &  47 $\pm$  20 & $<$0.05 & -- & $<$-0.1 $\pm$  0.5 \\
J182957.85+011251.1 & N & 0.05 $\pm$ 0.01 & $>$43 $\pm$  19 & $<$0.08 & -- & $<$ 0.8 $\pm$  0.5 \\
J182957.89+011246.0 & N & 3.20 $\pm$ 0.24 &  65 $\pm$   4 & 3.32 $\pm$ 0.41 &  62 $\pm$   8 &  0.1 $\pm$  0.2 \\
J182959.55+011158.1 & N & 0.07 $\pm$ 0.01 & $>$48 $\pm$  12 & $<$0.05 & -- & $<$-0.9 $\pm$  0.4 \\
J182959.94+011311.3 & N & 0.12 $\pm$ 0.02 &  42 $\pm$  20 & 0.10 $\pm$ 0.02 &  18 $\pm$  30 & -0.3 $\pm$  0.4 \\
J183000.65+011340.0 & N & 0.16 $\pm$ 0.02 & $>$82 $\pm$   3 & 0.10 $\pm$ 0.02 & $>$75 $\pm$   8 & -0.9 $\pm$  0.4 \\
J183001.24+010205.4 & N & 0.27 $\pm$ 0.04 & Extended & $<$0.06 & -- & $<$-3.1 $\pm$  0.3 \\
J183002.42+012405.6 & N & 0.25 $\pm$ 0.05 & Extended & $<$0.04 & -- & $<$-3.7 $\pm$  0.4 \\
J183002.67+012258.1 & Y & 0.15 $\pm$ 0.02 &  25 $\pm$  18 & 0.16 $\pm$ 0.04 &  14 $\pm$  31 &  0.1 $\pm$  0.3 \\
J183004.14+011239.7 & Y & 0.11 $\pm$ 0.01 &  28 $\pm$  19 & 0.07 $\pm$ 0.02 & $>$18 $\pm$  29 & -0.8 $\pm$  0.5 \\
J183004.62+012234.1 & N & 0.97 $\pm$ 0.11 &  49 $\pm$  10 & 1.06 $\pm$ 0.25 &  19 $\pm$  27 &  0.2 $\pm$  0.2 \\
J183004.65+011353.7 & Y & 0.07 $\pm$ 0.01 & $>$43 $\pm$  25 & $<$0.04 & -- & $<$-1.2 $\pm$  0.4 \\
J183004.98+012226.9 & Y & 0.30 $\pm$ 0.05 & Extended & $<$0.04 & -- & $<$-3.9 $\pm$  0.3 \\
J183007.29+010324.8 & Y & 0.12 $\pm$ 0.02 &  43 $\pm$  24 & 0.10 $\pm$ 0.02 &  34 $\pm$  25 & -0.3 $\pm$  0.4 \\
J183008.31+011519.1 & N & 0.13 $\pm$ 0.02 &  42 $\pm$  15 & 0.07 $\pm$ 0.02 & $>$ 6 $\pm$  34 & -1.2 $\pm$  0.5 \\
J183008.69+010631.3 & N & 3.18 $\pm$ 0.29 &  11 $\pm$  12 & 1.99 $\pm$ 0.35 &   5 $\pm$  24 & -0.9 $\pm$  0.1 \\
J183008.77+010257.7 & Y & 0.08 $\pm$ 0.01 & --A-- & 0.06 $\pm$ 0.02 & $>$30 $\pm$  27 & -0.5 $\pm$  0.6 \\
J183010.31+012345.2 & N & 0.77 $\pm$ 0.12 &  48 $\pm$  12 & -- & -- & -- \\
J183010.60+010320.7 & N & 0.19 $\pm$ 0.02 &  22 $\pm$  18 & 0.17 $\pm$ 0.03 &  36 $\pm$  22 & -0.3 $\pm$  0.3 \\
J183012.58+011226.8 & N & 0.07 $\pm$ 0.01 & $>$51 $\pm$  14 & $<$0.04 & -- & $<$-1.2 $\pm$  0.4 \\
J183014.25+010924.1 & Y & $<$0.03 & -- & 0.10 $\pm$ 0.03 & $>$22 $\pm$  29 & $>$ 2.1 $\pm$  0.6 \\
J183014.71+011629.6 & Y & 0.13 $\pm$ 0.02 &  40 $\pm$  15 & 0.10 $\pm$ 0.02 &  27 $\pm$  26 & -0.6 $\pm$  0.4 \\
J183015.53+012203.9 & Y & 0.20 $\pm$ 0.02 &  32 $\pm$  13 & 0.18 $\pm$ 0.03 &  43 $\pm$  20 & -0.2 $\pm$  0.3 \\
J183016.56+011304.3 & Y & 0.07 $\pm$ 0.01 & $>$36 $\pm$  15 & 0.11 $\pm$ 0.02 &  19 $\pm$  36 &  0.8 $\pm$  0.5 \\
J183016.74+010856.2 & Y & 0.27 $\pm$ 0.03 &  22 $\pm$  16 & 0.21 $\pm$ 0.05 &  22 $\pm$  28 & -0.5 $\pm$  0.2 \\
J183018.05+011819.2 & Y & $<$0.03 & -- & 0.09 $\pm$ 0.03 & $>$43 $\pm$  27 & $>$ 2.0 $\pm$  0.4 \\
J183022.13+011738.1* & Y & 0.06 $\pm$ 0.01 & --A-- & 0.05 $\pm$ 0.02 & $>$32 $\pm$  52 & -0.2 $\pm$  0.7 \\
J183024.87+011323.5 & N & 0.07 $\pm$ 0.01 & $>$20 $\pm$  23 & 0.08 $\pm$ 0.02 & $>$43 $\pm$  20 &  0.5 $\pm$  0.6 \\
J183025.10+012304.3 & N & 0.08 $\pm$ 0.01 &  37 $\pm$  26 & 0.10 $\pm$ 0.02 & $>$59 $\pm$  20 &  0.6 $\pm$  0.5 \\
J183031.05+011257.3 & N & 0.08 $\pm$ 0.02 &  43 $\pm$  21 & $<$0.05 & -- & $<$-1.0 $\pm$  0.4 \\
J183052.19+011915.5 & N & 0.10 $\pm$ 0.02 &  21 $\pm$  30 & -- & -- & -- \\
J183059.74+012511.7 & N & 0.96 $\pm$ 0.15 &  31 $\pm$  16 & -- & -- & -- \\
J183059.83+012516.5 & Y & 0.16 $\pm$ 0.03 &  34 $\pm$  20 & -- & -- & -- \\
J183059.86+012519.0 & Y & 0.20 $\pm$ 0.04 &  18 $\pm$  24 & -- & -- & -- \\
J183104.32+011309.0 & Y & 0.25 $\pm$ 0.05 &  42 $\pm$  17 & -- & -- & -- \\
\sidehead{Detected sources in the W40 region:}
J183044.11-020145.6 & N & 1.57 $\pm$ 0.20 &   9 $\pm$  16 & 1.27 $\pm$ 0.32 &  15 $\pm$  31 & -0.4 $\pm$  0.2 \\
J183023.27-020731.4 & Y & 0.25 $\pm$ 0.06 &  31 $\pm$  24 & 0.12 $\pm$ 0.06 &  36 $\pm$  48 & -1.5 $\pm$  0.4 \\
J183101.07-021136.8 & Y & 0.19 $\pm$ 0.03 &   6 $\pm$  20 & 0.14 $\pm$ 0.03 &  10 $\pm$  34 & -0.6 $\pm$  0.4 \\
J183102.25-015658.3 & Y & 3.25 $\pm$ 0.37 &   4 $\pm$  15 & 1.32 $\pm$ 0.31 &   0 $\pm$  33 & -1.8 $\pm$  0.2 \\
J183102.41-015706.2 & Y & 2.73 $\pm$ 0.33 & Extended & 0.96 $\pm$ 0.22 &   2 $\pm$  32 & -2.1 $\pm$  0.2 \\
J183102.94-015917.0 & Y & 0.21 $\pm$ 0.03 &  25 $\pm$  16 & 0.26 $\pm$ 0.06 &  50 $\pm$  18 &  0.4 $\pm$  0.3 \\
J183105.01-020247.6 & N & 0.09 $\pm$ 0.02 & $>$56 $\pm$  10 & 0.09 $\pm$ 0.02 & $>$41 $\pm$  22 & -0.0 $\pm$  0.6 \\
J183109.41-015442.1 & Y & 0.22 $\pm$ 0.03 &   3 $\pm$  24 & 0.16 $\pm$ 0.04 & Extended & -0.7 $\pm$  0.4 \\
J183113.05-021012.1 & Y & 4.70 $\pm$ 0.40 & Extended & 1.87 $\pm$ 0.22 & Extended & -1.9 $\pm$  0.3 \\
J183113.23-021011.1 & Y & 20.78 $\pm$ 1.22 & Extended & 11.50 $\pm$ 0.95 & Extended & -1.2 $\pm$  0.2 \\
J183114.31-020852.1 & Y & 0.13 $\pm$ 0.02 &  45 $\pm$  14 & 0.09 $\pm$ 0.02 &  12 $\pm$  36 & -0.7 $\pm$  0.5 \\
J183114.82-020350.1 & N & 0.36 $\pm$ 0.04 &  44 $\pm$   9 & 0.42 $\pm$ 0.07 &  50 $\pm$  13 &  0.3 $\pm$  0.3 \\
J183115.28-020415.2 & N & 0.80 $\pm$ 0.07 &   5 $\pm$  12 & 0.44 $\pm$ 0.07 &   1 $\pm$  22 & -1.2 $\pm$  0.2 \\
J183118.68-015455.9 & Y & 0.68 $\pm$ 0.07 &  11 $\pm$  14 & 0.50 $\pm$ 0.10 &   2 $\pm$  27 & -0.6 $\pm$  0.2 \\
J183119.86-020006.1 & Y & 0.19 $\pm$ 0.03 &  19 $\pm$  16 & 0.23 $\pm$ 0.05 &  15 $\pm$  25 &  0.3 $\pm$  0.3 \\
J183120.65-020943.6 & N & 0.09 $\pm$ 0.02 & $>$44 $\pm$  12 & 0.13 $\pm$ 0.03 &   9 $\pm$  33 &  0.7 $\pm$  0.5 \\
J183122.32-020619.6 & N & 0.67 $\pm$ 0.07 &  91 $\pm$   2 & 1.13 $\pm$ 0.22 &  96 $\pm$   5 &  1.1 $\pm$  0.2 \\
J183122.81-020930.7 & N & 0.09 $\pm$ 0.02 &   7 $\pm$  35 & 0.11 $\pm$ 0.03 & $>$25 $\pm$  23 &  0.4 $\pm$  0.5 \\
J183123.62-020535.8 & N & 3.52 $\pm$ 0.27 &  15 $\pm$   9 & 3.32 $\pm$ 0.45 &  16 $\pm$  16 & -0.1 $\pm$  0.1 \\
J183124.56-020231.9 & N & 0.07 $\pm$ 0.02 & $>$21 $\pm$  28 & $<$0.06 & -- & $<$-0.3 $\pm$  0.5 \\
J183125.77-015506.7 & Y & 2.90 $\pm$ 0.38 &  27 $\pm$  13 & 3.05 $\pm$ 0.83 &   5 $\pm$  37 &  0.1 $\pm$  0.2 \\
J183126.02-020517.0 & N & 1.00 $\pm$ 0.07 &  52 $\pm$   5 & 0.75 $\pm$ 0.09 &  61 $\pm$   9 & -0.6 $\pm$  0.2 \\
J183126.27-020630.7 & Y & 0.25 $\pm$ 0.06 & $>$39 $\pm$  10 & $<$0.11 & -- & $<$-1.7 $\pm$  0.5 \\
J183126.85-021042.4 & Y & 0.22 $\pm$ 0.03 &   9 $\pm$  19 & 0.14 $\pm$ 0.03 &  40 $\pm$  25 & -1.0 $\pm$  0.4 \\
J183127.30-020504.5 & N & 0.51 $\pm$ 0.05 &  20 $\pm$  11 & 0.45 $\pm$ 0.07 &  36 $\pm$  14 & -0.2 $\pm$  0.3 \\
J183127.45-020512.0 & N & 2.71 $\pm$ 0.19 &  96 $\pm$   1 & 3.17 $\pm$ 0.34 &  97 $\pm$   5 &  0.3 $\pm$  0.2 \\
J183127.64-020513.5 & N & 0.88 $\pm$ 0.08 & Extended & 0.65 $\pm$ 0.11 & Extended & -0.6 $\pm$  0.3 \\
J183127.65-020509.7 & N & 0.74 $\pm$ 0.05 &  19 $\pm$   9 & 0.77 $\pm$ 0.08 &  13 $\pm$  14 &  0.1 $\pm$  0.2 \\
J183127.67-020519.7 & N & 3.45 $\pm$ 0.24 &   6 $\pm$   9 & 3.49 $\pm$ 0.40 &   6 $\pm$  15 &  0.0 $\pm$  0.2 \\
J183127.78-020449.5 & N & 0.12 $\pm$ 0.02 &  23 $\pm$  22 & 0.11 $\pm$ 0.02 &   5 $\pm$  42 & -0.2 $\pm$  0.5 \\
J183127.78-020823.0 & Y & 0.33 $\pm$ 0.03 &   6 $\pm$  14 & 0.23 $\pm$ 0.02 &  14 $\pm$  15 & -0.7 $\pm$  0.3 \\
J183127.80-020521.9 & N & 1.47 $\pm$ 0.12 &  17 $\pm$   9 & 1.24 $\pm$ 0.15 &   3 $\pm$  17 & -0.3 $\pm$  0.2 \\
J183128.01-020517.9 & N & 0.61 $\pm$ 0.06 &  27 $\pm$  10 & 0.50 $\pm$ 0.07 &  18 $\pm$  16 & -0.4 $\pm$  0.2 \\
J183128.65-020529.8 & N & 6.76 $\pm$ 0.55 & Extended & 4.79 $\pm$ 0.71 & Extended & -0.7 $\pm$  0.2 \\
J183128.67-020522.2 & N & 0.31 $\pm$ 0.04 &  17 $\pm$  14 & 0.25 $\pm$ 0.04 &   3 $\pm$  22 & -0.4 $\pm$  0.3 \\
J183129.41-020541.1 & N & 0.18 $\pm$ 0.03 &   2 $\pm$  31 & 0.12 $\pm$ 0.03 & --A-- & -0.8 $\pm$  0.5 \\
J183130.54-020530.7 & N & 0.26 $\pm$ 0.05 &  18 $\pm$  27 & $<$0.07 & -- & $<$-2.8 $\pm$  0.4 \\
J183148.61-020700.7 & N & 0.08 $\pm$ 0.02 & $>$41 $\pm$  16 & $<$0.05 & -- & $<$-0.7 $\pm$  0.4 \\
J183153.39-020959.9 & N & 0.13 $\pm$ 0.02 &  36 $\pm$  20 & 0.08 $\pm$ 0.03 & $>$ 6 $\pm$  44 & -0.9 $\pm$  0.5 \\
J183201.69-020243.2 & Y & 0.71 $\pm$ 0.13 &   4 $\pm$  25 & -- & -- & -- \\
J183201.76-021012.1 & Y & 0.15 $\pm$ 0.03 &  21 $\pm$  27 & -- & -- & -- \\
\sidehead{Detected sources in the Serpens South cluster:}
J182933.58-014530.4 & Y & 0.24 $\pm$ 0.06 & $>$32 $\pm$  24 & -- & -- & -- \\
J182940.34-015127.9 & N & $<$0.08 & -- & 0.11 $\pm$ 0.03 & --A-- & $>$ 0.6 $\pm$  0.3 \\
J182952.73-015159.9 & Y & 0.23 $\pm$ 0.03 &  25 $\pm$  25 & 0.21 $\pm$ 0.04 &  40 $\pm$  26 & -0.2 $\pm$  0.3 \\
J183003.72-014944.3 & Y & 0.12 $\pm$ 0.03 & $>$45 $\pm$  19 & -- & -- & -- \\
J183004.81-020227.8 & N & $<$0.06 & -- & 0.08 $\pm$ 0.02 & --A-- & $>$ 0.7 $\pm$  0.5 \\
J183009.68-020032.7 & Y & 0.57 $\pm$ 0.06 &  43 $\pm$  10 & 0.64 $\pm$ 0.12 &  19 $\pm$  23 &  0.2 $\pm$  0.2 \\
J183018.69-020249.8 & Y & 0.20 $\pm$ 0.04 & $>$18 $\pm$  19 & $<$0.09 & -- & $<$-1.5 $\pm$  0.3 \\
J183025.24-021051.9 & N & 0.10 $\pm$ 0.03 & --A-- & $<$0.10 & -- & $<$-0.0 $\pm$  0.4 \\
J183031.68-020954.3 & Y & 0.25 $\pm$ 0.05 & $>$47 $\pm$  13 & -- & -- & -- \\
J183037.08-021503.3 & Y & 1.06 $\pm$ 0.18 &   7 $\pm$  23 & -- & -- & -- \\
J183038.25-021100.4 & Y & 0.26 $\pm$ 0.07 & $>$ 8 $\pm$  23 & -- & -- & -- \\
%%%%%
\enddata
\tablecomments{Flux densities are measured in the combined epochs images.
The quoted flux errors comprise the statistical error provided by IMFIT, the 5\% absolute flux uncertainty, and the uncertainty induced by the pointing error of the VLA primary beam. 
The asterisk indicates a source without reported counterparts detected with an integrated flux $< 5\sigma$, but with 
a peak flux $\geq 5\sigma$. 
The \emph{A} annotation indicates a source not detected at three times the noise level on individual epochs, but detected on the
image of the concatenated epochs.}
\tablenotetext{1}{Y=source without reported counterparts at any frequency. N=source with known counterpart.}
%\tablenotetext{a}{The first quoted errors correspond to the statistical error provided by IMFIT, while the second error reflects the 5\% absolute flux uncertainty and the uncertainty induced by the pointing error of the VL--A-- primary beam.}
%\tablenotetext{A}{Source not detected at three times the noise level on individual epochs, but detected on the
%image of the concatenated epochs.}
%\tablenotetext{B}{Source without reported counterparts detected with an integrated flux $< 5$-$\sigma$ but with 
%a peak flux $\geq 5$-$\sigma$.}
\label{tab:sources_sc}
\end{deluxetable}

\begin{deluxetable}{lccccccc}
\tabletypesize{\scriptsize}
\tablewidth{0pt}
\tablecolumns{8}
\tablecaption{Radio sources with known counterparts }
\tablehead{           & Other         &         & \multicolumn{3}{c}{Infrared\tablenotemark{b}} &       & Object\\
\colhead{GBS-VLA name} & \colhead{names} &\colhead{X-ray\tablenotemark{a}} & \colhead{SST} & \colhead{2M}&\colhead{WISE}&
\colhead{Radio\tablenotemark{c}}&\colhead{type}\\}
\startdata
%%%%%
\sidehead{Counterparts of the Serpens molecular cloud sources:}
J182851.30+010908.6 & -- &-- &-- &-- &-- & NVSS 182851+010908 &--\\
J182901.40+010434.7 & -- &-- &-- &-- &-- & NVSS 182901+010436 &--\\
J182905.07+012309.0 & --                   & --           & Y & -- & -- & --                             & -- \\
J182906.84+011742.7 & --                   & --           & Y & -- & -- & --                             & -- \\
J182907.62+012125.1 & --                   & --           & Y & -- & -- & --                             & -- \\
J182910.17+012559.5 & --                   & --           & Y & -- & -- & --                             & -- \\
J182913.17+010906.4 & --                   & --           & Y & -- & -- & --                             & -- \\
J182913.79+010738.6 & --                   & --           & Y & -- & -- & --                             & -- \\
J182916.11+010437.5 & --                   & --           & Y & -- & -- & --                             & -- \\
J182918.23+011757.7 & --                   & --           & Y & -- & -- & --                             & -- \\
J182926.71+012342.1 & --                   & --           & Y & -- & -- & --                             & -- \\
J182928.02+011156.5 & --                   & --           & Y & -- & -- & --                             & -- \\
J182928.28+011205.7 & -- &-- &-- &-- &-- & NVSS 182928+011203 &-- \\
J182929.78+012158.1 & --                   & GFM 6    & --  & -- & -- & --                           & YSO\\
J182930.71+010048.3 & PMN J1829+0101      &       --       & -- & -- & -- & NVSS 182930+010048  &    --  \\
J182933.07+011716.3 & SVS76 Ser 14 & GFM 11  &Y &Y  & Y&               --               & YSO \\
J182934.32+011513.9 & --                   & --           & Y & -- & -- & DCE08-210 5          & -- \\
J182935.02+011503.2 & DCE08-210 5         & --            & Y&   -- &-- &  NVSS 182934+011504       & --   \\
J182935.11+011503.6 & -- &-- &-- &-- &-- &  DCE08-210 5 &--\\
J182936.50+012317.0 & --                   & --           & Y & -- &-- & --                             & -- \\
J182937.76+010314.6 & -- &-- &-- &-- &-- & NVSS 182937+010316 &-- \\
J182944.07+011921.1 &  DCE08-210 6   & --           & -- &  -- & --& NVSS 182944+011920  &   --    \\
J182948.83+010647.4 & --                   & --           & Y & -- & -- & --                             & -- \\
J182949.79+011520.4 &       SERPENS SMM 1a & --           & -- & --  & --&  DCE08-210 1    &   --   \\
J182951.04+011533.8 & --                    & --           & -- & --  & -- & ETC 8                    & --     \\ 
J182951.17+011640.4 & V371 Ser      &  GFM 30   & Y &  -- & Y&  ETC 9    &  YSO   \\
J182951.22+012132.0 & -- &-- &-- &-- &-- & NVSS 182951+012131 &--\\
J182952.22+011547.4 & Serpens SMM 10 IR &   -- & Y & --  & Y& ETC 10                  & YSO \\
J182953.99+011229.5 & --                   &  --           & Y & --  & --&        --          &   --     \\
J182954.31+010309.6 & --                  & --           & Y & -- & -- & --                             & -- \\
J182956.96+011247.6 &     EC92 84        & GFM 44      & Y & Y & -- & ETC 14                       & YSO \\
J182957.60+011300.2 &  EES2009 Ser-emb 22  &  GFM 46 & Y & Y & --& ETC 15 & YSO \\
J182957.85+011251.1 &  EES2009 Ser-emb 23  &  GFM 53 & Y & Y & --& ETC 17 & YSO \\
J182957.89+011246.0 & EC92 95            & GFM 54  & Y & Y  & --&  DCE08-215 8       &   YSO \\
J182959.55+011158.1 &  EES2009 Ser-emb 24  &  GFM 60 & Y & Y & --&   --    & YSO \\
J182959.94+011311.3 & EES2009 Ser-emb 19&    --       & Y & --  &Y&        --                     & YSO \\
J183000.65+011340.0 & CK 6              & GFM 65   & Y & Y  &Y& ETC 20                  & YSO \\
J183001.24+010205.4 & -- &-- &-- &-- &-- & NVSS 183001+010204 &--\\
J183002.42+012405.6 &     --               &    --          & Y & --  &Y &        --                    &   --    \\
J183004.62+012234.1 & P2003 J183004.7+012232 & GFM 70   & -- &  -- &-- &         --                   &   --    \\
J183008.31+011519.1 & --                   & --           & Y & -- & -- & --                             & -- \\
J183008.69+010631.3 & -- &-- &-- &-- &-- & NVSS 183008+010634 &--\\ 
J183010.31+012345.2 & --                   & --           & Y & -- & -- & --                             & -- \\
J183010.60+010320.7 & --                   & --           & Y & -- & -- & --                             & -- \\
J183012.58+011226.8 &     --               & GFM 81   & Y &  -- &-- &         --                   &  --     \\
J183024.87+011323.5 & HD 170634   &   --           & Y &  Y &Y&          --                   & B7V  \\
J183025.10+012304.3 & --                   & --           & Y & -- & -- & --                             & -- \\
J183031.05+011257.3 & --                   & --           & Y & -- & -- & --                             & -- \\
J183052.19+011915.5 & --                   & --           & Y & -- & -- & --                             & -- \\
J183059.74+012511.7 & -- &-- &-- &-- &-- & NVSS 183059+012512 &--\\
\sidehead{Counterparts of the W40 region sources:}
J183044.11-020145.6 &     --             & --            & -- & Y & -- & --                        & -- \\
J183105.01-020247.6 & CXOW40 J183105.02-020247.5 & KGF 18  & --& Y & --&-- & YSO \\
J183114.82-020350.1 & W 40 IRS 5   & KGF 36   & -- & Y & -- & RRR W40-VLA 1 & B1V \\
J183115.28-020415.2 & CXOW40 J183115.30-020415.2 & KGF 38   & -- & -- & -- & RRR W40-VLA 2  & -- \\
J183120.65-020943.6 & CXOW40 J183120.65-020944.1 & KGF 71  & --& Y & --&-- & YSO \\
J183122.32-020619.6 & CXOW40 J183122.32-020619.5 & KGF 82   & -- & Y & -- & RRR W40-VLA 3  & YSO \\
J183122.81-020930.7 & CXOW40 J183122.82-020930.5 & KGF 88  & --& Y & --&-- & YSO \\
J183123.62-020535.8 & CXOW40 J183123.62-020535.7 & KGF 97   & -- & Y & -- & RRR W40-VLA 5  & YSO \\
J183124.56-020231.9 & CXOW40 J183124.57-020231.9 & KGF 102 & --& Y & --&-- & YSO \\
J183126.02-020517.0 & W 40 IRS 1c & KGF 122 & -- & Y & Y & RRR W40-VLA 8  & YSO  \\
J183127.30-020504.5 &    --              &   --          & -- & Y  & -- & RRR W40-VLA 9  & YSO? \\
J183127.45-020512.0 & CXOW40 J183127.46-020511.9 & KGF 133 & -- & Y & -- & RRR W40-VLA 10 & YSO? \\
J183127.64-020513.5 & CXOW40 J183127.64-020513.5 & KGF 136 & -- & Y & -- & RRR W40-VLA 12 & gyrosynchrotron \\
 &                                               &         &    &   &    &                & source?\\
J183127.65-020509.7 & W 40 IRS 1d & KGF 138 & -- &Y & -- & RRR W40-VLA 13 & YSO \\
J183127.67-020519.7 &    --              &    --         & -- & Y & -- & RRR W40-VLA 14 & gyrosynchrotron \\
                    &                    &               &    &   &    &                &  source? \\
J183127.78-020449.5 & CXOW40 J183127.78-020449.5 & KGF 139 & --& Y & --&-- & YSO \\
J183127.80-020521.9 &  W 40 IRS 1a N  &       --         & -- & Y & -- & RRR W40-VLA 15  & YSO \\
J183128.01-020517.9 & CXOW40 J183128.01-020517.1 & KGF 144 & -- & Y & -- & RRR W40-VLA 16 & YSO \\
J183128.65-020529.8 & W 40 IRS 1b & KGF 145 & -- & Y & -- & RRR W40-VLA 18 & YSO \\
J183128.67-020522.2 &    --               &   --          &-- & -- & -- & RRR W40-VLA 19 & shock front? \\
J183129.41-020541.1 & CXOW40 J183129.45-020541.2 & KGF 153 & --& Y & --&-- & YSO \\
J183130.54-020530.7 & CXOW40 J183130.56-020530.6 & KGF 162 & --& Y & --&-- & YSO \\
J183148.61-020700.7 & CXOW40 J183148.64-020755.5 & KGF 220 & --& Y & Y &-- & YSO \\
J183153.39-020959.9 &    --               &     --        & -- & Y & Y & --                         & -- \\
\sidehead{Counterparts of the Serpens South sources:}
J182940.34-015127.9 &  --                  &     --         & Y & -- & -- & -- & YSO \\
J183004.81-020227.8 &  --                  &     --         & Y & -- & -- & -- & YSO \\
J183025.24-021051.9 &  --                  &     --         & Y & -- & -- & -- & YSO  \\
%
% 
%%%%%
%
\enddata
\tablenotetext{a}{GFM = Giardino et al.\ (2007); KGF = Kuhn et al.\ (2010)}
\tablenotetext{b}{SST = Evans et al.\ (2009), c2d-GB clouds catalog (Dunham et al. 2013 and Allen et al. (in preparation))
; 2M = Cutri et al.\ (2003) and WISE = Cutri
et al\ (2012).}
%\tablenotetext{c}{For GBDS VLA J162722.96-242236.6, 2MASS and SST data from Marsh et al.\ (2010).}
\tablenotetext{c}{ETC = Eiroa et al.\ (2005); 
NVSS = Condon et al. (1998); DCE08 = Scaife et al. (2012);
%PMN = Griffith et al.\ (1995); 
%SB86 = Snell \& Bally (1986);
%B96 = Bontemps\ (1996); 
RRR = Rodr{\'{\i}}guez et al.\ (2010).}%%%%%
\label{tab:sources_count1}
\end{deluxetable}

\begin{deluxetable}{ccc}
\tabletypesize{\scriptsize}
\tablewidth{0pt}
\tablecolumns{3}
\tablecaption{Sources detected in circular polarization \label{tab:cir_pol}}

\tablehead{
\colhead{GBS-VLA name} & \multicolumn{2}{c}{Degree of circular polarization} \\
\colhead{} & \colhead{4.5 GHz (\%)} & \colhead{7.5 GHz (\%)} \\ 
}
\startdata
%%%%%
J183113.23-021011.1 & 1 (L)& 1 (R)\\
J183123.62-020535.8 & 5 (R)& 8 (R)\\

\enddata
\tablecomments{The letters indicate
left (L) or right (R) circular polarization.}
\end{deluxetable}

\begin{deluxetable}{lrrr}
\tabletypesize{\scriptsize}
\tablewidth{0pt}
\tablecolumns{4}
\tablecaption{Young stellar object candidates based just on their radio properties \label{tab:ysoC}}
\tablehead{ GBS-VLA name & \colhead{Variability [4.5 GHz]} & \colhead{Variability [7.5 GHz]} & Spectral index \\
\colhead{} & (\%) & (\%) & \colhead{} \\}
\startdata
%%%%%
\cutinhead{YSO candidates in the Serpens Molecular cloud}
J182854.46+011823.7 &  44 $\pm$   9 &  17 $\pm$  24 &  1.2 $\pm$  0.1 \\
J182910.17+012559.5 &  71 $\pm$   8 & -- & -- \\
J182913.79+010738.6 &  36 $\pm$  25 & $>$26 $\pm$  20 &  0.9 $\pm$  0.5 \\
J182916.11+010437.5 &  41 $\pm$  16 &  52 $\pm$  19 &  0.4 $\pm$  0.2 \\
J182932.21+012104.6 &  52 $\pm$  15 & $>$20 $\pm$  27 & -0.8 $\pm$  0.5 \\
J182948.83+010647.4 &  29 $\pm$  13 &   6 $\pm$  23 &  0.8 $\pm$  0.2 \\
J182948.92+011523.8 & $>$62 $\pm$  10 & -- & $<$-0.5 $\pm$  0.5 \\
J182949.54+011523.8 &  76 $\pm$   7 & -- & $<$-2.5 $\pm$  0.3 \\
J182949.60+011522.9 &  71 $\pm$   8 &  64 $\pm$  14 & -0.5 $\pm$  0.4 \\
J182950.34+011515.3 &  11 $\pm$  20 & $>$61 $\pm$  18 & -1.0 $\pm$  0.4 \\
J182954.30+012011.2 & $>$75 $\pm$   5 & -- & $<$-2.3 $\pm$  0.3 \\
J183012.58+011226.8 & $>$51 $\pm$  14 & -- & $<$-1.2 $\pm$  0.4 \\
J183014.25+010924.1 & -- & $>$22 $\pm$  29 & $>$ 2.1 $\pm$  0.6 \\
J183016.56+011304.3 & $>$36 $\pm$  15 &  19 $\pm$  36 &  0.8 $\pm$  0.5 \\
J183018.05+011819.2 & -- & $>$43 $\pm$  27 & $>$ 2.0 $\pm$  0.4 \\
\cutinhead{YSO candidates in the W40 region}
J183127.45-020512.0 &  96 $\pm$   1 &  97 $\pm$   5 &  0.3 $\pm$  0.2 \\
%%%%%
\enddata
%\tablenotetext{a}{}
%\tablenotetext{b}{}
\end{deluxetable}

\begin{deluxetable}{lcccccc}
\tabletypesize{\scriptsize}
\tablewidth{0pt}
\tablecolumns{7}
\tablecaption{Young stellar objects detected in the radio observations \label{tab:yso}}
\tablehead{           & Spectral       &  SED  & High\tablenotemark{b} & & & \\
\colhead{GBS-VLA name} & \colhead{type}& \colhead{classification\tablenotemark{\bf{a}}}& \colhead{variability}  & 
\colhead{$\alpha$\tablenotemark{b}}&\colhead{X-ray} &\colhead{Reference\tablenotemark{c}}\\}
\startdata
%%%%%
\cutinhead{YSOs in the Serpens molecular cloud}
J182929.78+012158.1 & --& Class II & -- & N & Y & 1 \\
J182933.07+011716.3 & G2.5 & Class III  & Y & P & Y & 1, 5\\
J182951.17+011640.4 & -- & Class I & Y & N & Y & 1, 7\\
J182952.22+011547.4 & --& Class I & N & N & N & 2, 7\\
J182956.96+011247.6 & M3.0 & Class II & Y & P & Y & 1, 5 \\
J182957.60+011300.2 & --& Class I & N & F & Y & 2, 7 \\  
J182957.85+011251.1 & --& Class I & N & P & Y & 2, 7 \\
J182957.89+011246.0 & K1.0 & P-HAeBe  & Y & F & Y & 6 \\
J182959.55+011158.1 & --& Class I & N & N & Y & 2, 7 \\
J182959.94+011311.3 & --& Class I & N & N & N & 2, 7\\
J183000.65+011340.0 & M0.5 & Class III  & Y & N & Y & 1, 5\\
%w40::
%\hline
\cutinhead{YSOs in the W40 region}
%\sidehead{YSOs in the W40 region}
%\hline
J183105.01$-$020247.6 & --& Class III& Y & F & Y & 3\\ 
J183120.65$-$020943.6 & --& Class III& N & P & Y & 3\\ 
J183122.32$-$020619.6 & --& Class III& Y & P & Y & 3 \\ %N
J183122.81$-$020930.7 & --& Class III& N & P & Y & 3\\
J183123.62$-$020535.8 & --& Class III& N & F & Y & 3\\ %N
J183124.56$-$020231.9 & --& Class III& N & N & Y & 3\\
J183126.02$-$020517.0 & --& Class II   & Y & N & Y & 4\\ 
J183127.65$-$020509.7 & --& HAeBe?\tablenotemark{d}& N & F & Y & 3\\ %N
J183127.78$-$020449.5 & --& Class III& N & F & Y & 3\\ 
J183127.80$-$020521.9 & --& HAeBe & N & N & N & 4\\ 
J183128.01$-$020517.9 & --& -- & N & N & Y & 3 \\
J183128.65$-$020529.8 & --& Class II & Extended & N & Y & 4\\
J183129.41$-$020541.1 & --& Class III& N & N & Y & 3\\ 
J183130.54$-$020530.7 & --& Class III& N & N & Y & 3\\
J183148.61$-$020700.7 & --& Class III& N & N & Y & 3\\
\cutinhead{YSOs in Serpens South}
J182940.34$-$015127.9 & -- & Flat     & --& P &N& 7 \\
J183004.81$-$020227.8 & -- & Class I  & --& P &N& 7 \\
J183025.24$-$021051.9 & -- & Class II & --& F &N& 7 \\
%\sidehead{a}
%\sidehead{b}
%%%%%
  % \label{tab:yso}
\enddata
\tablenotetext{a}{This classification was taken from the literature; references are given in column 7.}
\tablenotetext{b}{High variability\ = Y when the flux variability is $\apprge 50\%$ at a 3$\sigma$ level
 in at least one frequency; N when the variability is $< 50\%$ at both frequencies. $\alpha$ 
refers to the spectral index, and is given as P (for positive) when it is higher than 0.2; F (for flat) 
when it is between --0.2 and $+$0.2; and N (for negative) when is is lower than --0.2. X-ray\ = Y when 
there is an X-ray flux reported in literature, N when it is not.}
\tablenotetext{c}{1= Giardino et al.\ (2007); 2 = Enoch et al. (2009); 3 = Kuhn et al.\ (2010);
4 = Shuping et al.\ (2012); 5 = Winston et al.\ (2010); 6 = Preibisch (1999); (7) = c2d-GB clouds catalog 
(Dunham et al. 2013 and Allen et al. (in preparation))}
\tablenotetext{d}{This source has a mass of $\sim 4~\rm{M}_{\odot}$ and no $K_s$-band-excess. 
We therefore consider it as an HAeBe candidate.}
\end{deluxetable}

\begin{deluxetable}{ccccc}
\tabletypesize{\scriptsize}
\tablewidth{0pt}
\tablecolumns{7}
\tablecaption{Proper motions of some radio sources in W40 \label{tab:prop_mov}}
%\tablehead{  Source  & \multicolumn{3}{c}{260 pc}    &   \multicolumn{3}{c}{415 pc} \\
\tablehead{
\colhead{GBS-VLA name} & \colhead{RRR W40-VLA} & \colhead{$\mu_\alpha \cos (\delta)$}& \colhead{$\mu_\delta$}& \colhead{$\mu_{\rm total}$} \\
\colhead{}& \colhead{number$^a$} & \colhead{(mas yr$^{-1}$)}  &\colhead{ (mas yr$^{-1}$)}  & \colhead{(mas yr$^{-1}$)}\\ 
}
\startdata
%%%%%
J183122.32-020619.6 & 3 & 5.13 $\pm$ 1.99 & -12.1 $\pm$ 2.15 & 13.14 \\
J183123.62-020535.8 & 5  & -2.43 $\pm$ 0.4 & -6.83 $\pm$ 0.42 & 7.25 \\
J183126.02-020517.0 & 8  & -7.58 $\pm$ 2.41 & -12.27 $\pm$ 2.59 & 14.43 \\
J183127.30-020504.5 & 9  & -7.54 $\pm$ 3.96 & -10.32 $\pm$ 4.22 & 12.78 \\
J183127.67-020519.7 & 14 & -9.17 $\pm$ 0.64 & -7.77 $\pm$ 0.67 & 12.02 \\
J183127.80-020521.9 & 15 & -6.94 $\pm$ 1.0 & -9.95 $\pm$ 1.06 & 12.13 \\
J183128.01-020517.9 & 16 & -6.79 $\pm$ 2.28 & -7.56 $\pm$ 2.43 & 10.16 \\
J183128.65-020529.8 & 18 & -8.29 $\pm$ 2.04 & -12.24 $\pm$ 2.14 & 14.78 \\
J183128.67-020522.2 & 19 & -14.97 $\pm$ 3.8 & -10.38 $\pm$ 4.14 & 18.22 \\
%\sidehead{a}
%\sidehead{b}
%%%%%
  % \label{tab:yso}
\enddata
\tablenotetext{a}{The labels in this column refer to the VLA source number in the catalog of  
Rodr\'iguez et al. (2010). }
\end{deluxetable}

%\begin{figure}[!ht]
%\includegraphics[width=1.1\textwidth,angle=0]{w402.eps}
%\caption{Same as Figure \ref{fig:w401}. }
%\label{fig:w402}
%\end{figure}

%%%%%%%%%%%%%%%%%%%%%%%%%%%%%%%%%%%%%%%%%%%%%%%%%%%%%%

\clearpage

\end{document}